\documentclass[preprint,12pt,authoryear]{elsarticle}
 \pdfoutput=1
\usepackage{amssymb}

\journal{Advances in Space Research}

\begin{document}

\begin{frontmatter}

%% Title, authors and addresses

%% use the tnoteref command within \title for footnotes;
%% use the tnotetext command for theassociated footnote;
%% use the fnref command within \author or \address for footnotes;
%% use the fntext command for theassociated footnote;
%% use the corref command within \author for corresponding author footnotes;
%% use the cortext command for theassociated footnote;
%% use the ead command for the email address,
%% and the form \ead[url] for the home page:
%% \title{Title\tnoteref{label1}}
%% \tnotetext[label1]{}
%% \author{Name\corref{cor1}\fnref{label2}}
%% \ead{email address}
%% \ead[url]{home page}
%% \fntext[label2]{}
%% \cortext[cor1]{}
%% \address{Address\fnref{label3}}
%% \fntext[label3]{}

\title{Sea state bias in altimetry measurements \\within the theory of similarity for wind-driven seas}

\author[label1,label2]{Sergei I. Badulin\corref{cor1}}
{\author[label1]{Vika G. Grigorieva}
\author[label1]{Pavel A. Shabanov}
\author[label1]{Vitali D. Sharmar}
\author[label1]{Ilya O. Karpov}}

\cortext[cor1]{Corresponding author: Sergei I. Badulin \ead{badulin.si@ocean.ru}}

\address[label1]{Shirshov Institute of Oceanology, Russian Academy of Sciences, 36, Nakhimovskii pr., 117997 Moscow}
\address[label2]{Skolkovo Institute of Science and Technology, Bolshoy Boulevard 30, bld. 1, 121205  Moscow, Russia}

\begin{abstract}
%% Text of abstract
   The theory of similarity for wind-driven seas is applied to the physical analysis of the problem of sea state bias (SSB) in altimetry measurements. Dimensionless wave steepness and pseudo-age derived from altimetry measurements are expected to provide {physically relevant and accurate enough}  SSB estimates. Analysis of Jason-1,2,3 and SARAL/AltiKa data within the approach shows the similarity and robustness of SSB distributions re-casted onto space of wave pseudo-age and steepness. This result is considered as a ground for developing a new parametric model of SSB and for analysis of underlying physical effects.

\end{abstract}

\begin{keyword}
sea state bias \sep satellite altimetry \sep wave steepness \sep pseudo-age of wind waves \sep similarity approach
%% keywords here, in the form: keyword \sep keyword
%% PACS codes here, in the form: \PACS code \sep code
%% MSC codes here, in the form: \MSC code \sep code
%% or \MSC[2008] code \sep code (2000 is the default)
\end{keyword}

\end{frontmatter}

% \linenumbers

\section[Introduction]{Introduction}
\label{sect1}

Sea state bias (SSB) in radar altimeter measurements is one of the most complex sources of the ranging errors. Being of the order of a few percents of significant wave height it is the largest remaining error in mean sea level estimates \citep{Gommen25yearon}.

Three main effects are assumed to be responsible for SSB. Electromagnetic bias (EMB) is associated with different reflectivity of wave crests and troughs and resulting in a permanent overestimation of satellite-to-surface range  measured by  altimeter. Skewness bias (SB) is connected with the inherently nonlinear dynamics of water waves. It makes wave surface to be  vertically asymmetric with flat troughs  and sharp crests and, thus, also leads to similar overestimation. Numerous instrumental and retracking effects are usually related to tracker bias (TB). All the above  effects are difficult to discriminate and describe within a consistent physical model which is relevant to the today altimetry needs.

Parametric models  are widely used when SSB is represented as a function of  geophysical predictors measured by altimeter itself or/and provided by sea wave  models \citep[e.g. WaveWatch III, see][]{WWIIImanual516}. The parameterization in terms of the key predictor, significant wave height, $H_s$ provides rather good fit of the SSB \citep[e.g.][p.46]{HandbookAltika2016} with the SSB coefficient $\alpha=SSB/H_s$ in the range of $2-5\%$ \citep{Gommen25yearon}. Further corrections refer to the wind speed $U_{10}$ derived from the normalized radar cross-section $\sigma_0$ (NRCS) and the  series in powers of $H_s$ and $U_{10}$ \citep[e.g.][]{Gaspar1994}. The parameterizations in terms of wind speed $U_{10}$ show, in particular, a pronounced regional dependence \citep{FuGlazman1991}  that can be explained mostly by the incompleteness of the set of predictors $H_s,\, U_{10}\, (\sigma_0)$ for  wave dynamics and, then, for the   SSB assessment. Using the characteristic wave period as an additional predictor shows some success  in the empirical models of SSB \citep{PiresEtal2016,PiresEtal2018}. However, the wave period, being a function of altimeter measured $H_s$ and $\sigma_0$ does not extend the number of independent physical predictors  only changing the ansatz of the parametric formula.

The role of wave steepness (wave surface slope) as a physical predictor of sea state bias has been realized a long time ago \citep{Longuet63,Srokosz86,GlazmanSrokosz1991} and verified in case studies within wave forecasting models and in situ wave measurements \citep{Millet2003,GommenBias2003}. The approach has not been applied to satellite data in the absence of wave steepness measurements by altimeters. A recent physical model of wave steepness from altimetry data \citep{Badulin2014,BadulinEtal2018} made this key parameter of wave dynamics readily available.

We present a physical analysis of SSB based on altimeter-derived wave steepness and pseudo-age. Wave steepness  is defined in terms of $H_s$ and spectral peak period $T_p$ \citep[cf.][]{GommenBias2003}
\begin{equation}\label{eq2}
  \mu=\frac{\pi H_s}{gT_p^2}.
\end{equation}
and can be estimated from along-track derivatives of $H_s$ as follows \citep{Badulin2014,BadulinEtal2018}
\begin{equation}\label{eq3}
  \mu=0.596|\nabla H_s|^{1/5}.
\end{equation}
In theoretical models of the EM scattering wave steepness is introduced as an integral  \citep[e.g.][]{GommenBias2003} through a range of contributing wave scales. However, the convergence of the integral requires  physically relevant high-frequency cutoff. Definition of Eq.~(\ref{eq2})  is free of such problems but, the method of assessment of $\mu$ with  Eq.~(\ref{eq3}) also puts certain limitations.

 In contrast to the  point-wise measured parameters $H_s$ and $\sigma_0$ wave steepness $\mu$ in Eq.~(\ref{eq3}) reflects a wave field evolution along the satellite track. On the one hand, Eq.~(\ref{eq3}) allows one to capture instantaneous sea state characteristics. On the other hand, the experimental estimates with Eq.~(\ref{eq3}) are vulnerable  to higher random errors of the gradient $\nabla H_s$ as compared with its point-wise counterparts $H_s,\,\sigma_0$. As recompense, the gradient operation achieves error minimization of $H_s$ measurements  and does not  correlate with  $H_s,\,\sigma_0$. Additionally, low exponent $1/5$ in Eq.~(\ref{eq3}) reduces the random errors of the resulting estimates of $\mu$.

 The simplest parametric model of SSB claims a proportionality to significant wave height
\begin{equation}\label{eq4}
  SSB/H_s=\alpha_l.
\end{equation}
The coefficient $\alpha_l$ being typically in a range of  $0.02-0.05$ ($2-5\%$) \citep{Gommen25yearon} provides a good reference but shows a dependence on chosen region and  altimeter mission.

Advanced physical model of  \citet{FuGlazman1991} relates SSB with the so-called pseudo wave age
\begin{equation}\label{eq5}
  \xi=\frac{gH_s}{U_{10}^2}.
\end{equation}
in a form of power-law dependence
\begin{equation}\label{eq6}
  \frac{SSB}{H_s}=A\left(\frac{\xi}{\xi_m}\right)^m
\end{equation}
Parameters  $A,\,m$ ($\xi_m$ -- `mean-over-globe' pseudo-age) in Eq.~(\ref{eq6}) also show a pronounced regional variations \citep[see Table 1 in][]{FuGlazman1991} and reflects insufficiency of the only altimeter-derived parameter to describe variability of SSB with required accuracy.

In \S~2 we  propose a consistent similarity approach for wind-driven seas \citep[e.g.][]{Kitai62} for the specific problem of sea state bias in altimetry measurements. The approach is based on altimeter-derived dimensionless wave steepness $\mu$ and pseudo-age $\xi$. The use of dimensionless variables instead of dimensional $H_s,\,U_{10}$ implies a more transparent physical analysis of the combined effects on SSB.

\S~3 is focused on the analysis of  satellite data. {Data of the TOPEX/Poseidon follow-on missions Jason-1,2,3 (Ku-band $\lambda=2.2$cm)  and of SARAL/AltiKa Ka-band radar ($\lambda=0.8$cm) shows good prospects of developing the parametric models based on $\xi,\,\mu$.}

Discussion outlines the results and perspectives of the proposed approach.

\section[Similarity approach]{Similarity approach for the  sea state bias in altimetry measurements}
\label{sect2}
%%%%%%%%%%%%%%
\subsection[Dimensional parameters]{Dimensionless parameters as physical predictors of the sea state bias}
 Assessment of sea state bias in altimetry is a  complex multidisciplinary problem.  Peculiarities of nonlinear water wave dynamics, electromagnetic scattering from the random sea surface, data assimilation and processing heavily affect the resulting ranging errors. Historically the SSB models rely upon altimeter-derived dimensional values of significant wave height $H_s$ and the normalized radar cross-section (NRCS) $\sigma_0$. Conversion of these measurable parameters into near-surface wind speed $U_{10}$ {allows for associating results with in-situ sea state  \citep[e.g.][]{Gommen2003,Quilfen2004,Mackay2008}. However, $U_{10}$ being estimated with parametric models contributes an additional error up to 20\%   \citep{ArdhuinFrontiers2018}.}

A consistent physical approach can be developed based on the well-elaborated similarity approach. By following this approach the number of  arguments of a physical problem can be reduced by a number of independent physical dimensions  \citep[the so-called $\Pi$-theorem, see][]{Barenblattbook79}. The non-dimensionalization appears to be a valuable research tool even (in our case, especially) if the form of the governing equations is still unknown.

Potentially, the similarity approach can help to  associate   the physical effects with underlying physical mechanisms.

Existing  models of SSB propose the following set of dimensional values:
\begin{enumerate}
  \item $g$ -- gravity acceleration  responsible for occurrence sea surface waves;
  \item $H_s$ -- significant wave height directly measured by altimeter;
  \item $U_{10}$ -- altimeter-derived wind speed that depends mostly on dimensionless NRCS $\sigma_0$ measured by altimeter.
\end{enumerate}
Two independent dimensions of the above set provides the only dimensionless parameter $\xi=gH_s/U_{10}^2$, the so-called pseudo-age of wind-driven waves. Based on physical variables $1-3$ the similarity approach gives a simple model of SSB \citep[cf. Eq.~(\ref{eq5}) and][]{FuGlazman1991}
\begin{equation}\label{eq7}
  \frac{SSB}{H_s}=F_1(\xi)
\end{equation}
Eq.~(\ref{eq7}) can be generalized by introducing the altimeter-derived  wave steepness $\mu$ \citep{Badulin2014,BadulinEtal2018}
\begin{equation}\label{eq8}
  \frac{SSB}{H_s}=F_2(\xi, \mu)
\end{equation}
Pseudo-age $\xi$ and wave steepness $\mu$ are naturally associated with wind-wave coupling and inherently nonlinear wave dynamics but they cannot completely describe SSB. However, accounting additional physical effects inevitably complicate the model.

As an example,  incorporating the well-known effect of radar wavelength  $\lambda_r$ on the SSB  \citep[e.g.][]{Melville1991,Walsh1991,GommenBias2003} results in the extension of Eqs.~(\ref{eq7},\ref{eq8})
\begin{equation}\label{eq9}
  \frac{SSB}{H_s}=F_3(\xi, \mu,\nu)
\end{equation}
with the third dimensionless argument $\nu$ that can be presented in two different ways
\begin{equation}\label{eq9a}
  \nu_1=\frac{ U_{10}^2}{g \lambda_r}; \qquad \nu_2=\frac{H_s}{\lambda_r}.
\end{equation}
The second form $\nu_2$ in Eq.~(\ref{eq9a}) is preferable because it uses the directly measured wave height and not the empirically estimated value of wind speed like in $\nu_1$. Additionally, being dependent on  square of wind speed $U_{10}$, the variable $\nu_1$ implies higher random errors than $\nu_2$.

 The three-parametric dependencies Eq.~(\ref{eq9}) are justified when comparing measurements of radars with different wavelengths (e.g. Ku-band $\lambda \approx 2.2$cm \emph{vs} Ka-band $\lambda\approx 0.8$cm). However, in our case when the altimeter wavelength is fixed, the two-parametric dependency Eq.~(\ref{eq8}) is enough. Note, that conventional counterparts also operate with two-dimensional parameterizations and can be formally reduced to the following form
 \begin{equation}\label{eq9b}
SSB=F_{conv}(\nu_1,\nu_2).
 \end{equation}
Such a representation can be considered a special case of the formal application of the similarity approach, or an expression of a completely different physical conception, when the main role is given to the processes of electromagnetic scattering ($\lambda_r$ and the associated electromagnetic bias EMB) while the role of wave dynamics (steepness $\mu$ and skewness bias SB) is belittled.

\subsection[Pseudo-age]{Pseudo-age of wind-driven waves: stages of wave growth and SSB effects}
%%%%%%%%%%%%%%%%%%%%%%%%%%
The first argument of $F_{2}$ in Eq.~(\ref{eq8}), pseudo-age $\xi$, is widely used in sea wave studies since \citet[][see Fig.7 therein]{SverdrupMunk1947}. Basic similarity theory of wind-driven waves \citep{Kitai62} predicts its dependence on alternative dimensionless parameter, wave age $a=C_p/U_{10}=gT_p/(2\pi U_{10})$ ($T_p$ being wave spectra peak period and $C_p$ is the corresponding wave phase speed). This dependence is not universal and can heavily depend on features of wind-wave coupling: stratification of wind flow, wind gustiness etc. \citep[see discussion in][]{Abdalla_Cavaleri2002,DonelanEtal2005,BBRZ2007}.

At specific stages of wave field evolution, the dependencies $\xi(a)$ follow power laws quite well. Exponents and pre-exponents of these dependencies have found their consistent treatment within the weak turbulence theory \citep[e.g.][and refs. therein]{Zakh2018} supported by solid experimental background \citep{Toba1972,Hass_ross_muller_sell76,ZakhZasl83b}. Wave age $a$ as a ratio of wave and wind characteristic speeds  remains a more effective parameter of physical analysis: $ a \lesssim 1$ is treated as a case of growing wind seas while $a \gtrsim 1$ is referred to swell. A more detailed scheme of wave evolution can be developed based on the results of \citet{GBB2011,BadulinGeog2019}.

In case of young seas  \citep[$a \ll 1$, generally $a < 1/3$, $\xi\sim a^{5/3}$, ][see Fig.1 and Table 1]{Hass_ross_muller_sell76,BadulinGeog2019}  short and steep wind waves of relatively small amplitudes provide a minor contribution into SSB through a  mechanism of SB (skewness bias). At the same time, high local wave steepness and the resulting white-capping can increase dramatically the effect of EMB (electromagnetic bias) both in terms of absolute and normalized value $SSB/H_s$.

Growing wind seas in the range of $1/3 \lesssim a \lesssim 1$, $\xi \sim a^{3/2}$ are associated with the so-called Toba regime \citep{Toba1972}. Increasing wave height amplifies absolute SB while wavelength growth (spectra downshift) partially suppresses this effect for normalized SSB.

At high ages (old waves $a>1$), wave growth and spectral downshifting is slowing down, then passing to hypothetical fully developed (matured) sea \citep{KomenHass84} or/and to swell. Skewness bias (SB), thus, is stabilizing or slowly decaying while the effect of EMB cannot be treated so straightforwardly. White-capping can be suppressed  at this stage as well as sea surface drag coefficient \citep{KudryavtsevMakin2011} and the effect of EMB can be reduced accordingly.

{On the one hand, the dependence of SSB on pseudo-age $\xi$ (age $a$) shows rich physics. On the other hand, this physics implies a non-trivial coupling of EMB and SB fractions of  SSB. The  theoretical and experimental issues of discrimination of these fractions in the context of sea state bias are discussed below.}

\subsection[Wave steepness]{Wave steepness: skewness bias of random weakly non-gaussian  sea }
%%%%%%%%%%%%%%%%%%%%%%%
In contrast to multiple physical effects associated with pseudo-age  $\xi$ wave steepness can be inherently  related to the effect of skewness bias. Analysis of EM scattering from a weakly non-gaussian random surface by \citet{GlazmanSrokosz1991} can be extended in view of parameterization Eq.~(\ref{eq8}).

The outcome of the analysis \citep[][]{Longuet63,Jackson1979,Srokosz86} for satellite altimetry is expressed by the following relationship \citep[see also][]{GommenBias2003}
\begin{equation}\label{eq10}
  SSB=\frac{1}{8}\left(\frac{\lambda_0}{3} + \lambda_1\right)H_s
\end{equation}
in notations of \citet{GlazmanSrokosz1991}. Here $H_s=4\langle \zeta^2 \rangle^{1/2}$ is defined in terms of sea surface elevation $\zeta$, $\lambda_0=\langle\zeta^3\rangle/\langle \zeta^2\rangle^{3/2}$  is the surface skewness, $\lambda_1$ was called `cross-skewness' by \citet{Srokosz86}
\begin{equation}\label{eq11}
  \lambda_1=\frac{\lambda_{120}+\lambda_{102}-2\lambda_{011}\lambda_{111}}{1-\lambda_{011}^2}
;\,
  \lambda_{mnp}=\frac{\mu_{mnp}}{\mu_{200}^{m/2} \mu_{020}^{n/2}\mu_{002}^{p/2}}; \, \mu_{mnp}=\langle \zeta^m\zeta^n\zeta^p \rangle.
\end{equation}
and later referred as `specular height' \citep{GlazmanSrokosz1991}. It can be shown \citep{Longuet63,BarrickLipa1985,Srokosz86} that both $\lambda_0$ and $\lambda_1$ may be expressed in terms of integrals of the two-dimensional wavenumber spectrum. This feature of Eq.~(\ref{eq10}) has been realized in estimates of SSB for simulated and observed sea wave spectra \citep{GommenBias2003} and developing a wind-speed based physical model of SSB by \citet{GlazmanSrokosz1991}.

In the latter case, explicit forms of spectra have been used that rely on classic solutions of the weak turbulence theory \citep{ZakhFil66,ZakhZasl83a,ZakhZasl83b} and on empirical dependencies of wave energy on wave age $a=C_p/U_{10}$. For unidirectional sea the corresponding algebra leads to explicit analytical expressions for $\lambda_0,\,\lambda_1$ and power-law dependence of SSB on wave age (pseudo-age) in the form of Eq.~(\ref{eq6}) \citep[see][sects.4--8]{GlazmanSrokosz1991}. Quite similar algebra can be realized for wave spectra written in the form \citep[see notations in ][]{GlazmanSrokosz1991}
\begin{equation}\label{eq13}
  E(k)=\beta g^2 k^{-5/2-n} k_p^{-1/2+n} \Phi\left(\frac{k}{k_p}\right).
\end{equation}
At exponent $n=0$ Eq.~(\ref{eq13}) is consistent with the classic direct cascade solution by \citet{ZakhFil66} with spectra tail asymptote $k^{-5/2}$ and a shape described by function $\Phi$ of dimensionless wavenumber $k/k_p$ ($k_p$ -- spectral peak wavenumber). The spectrum magnitude $\beta$ in Eq.~(\ref{eq13}), evidently, is related to wave steepness $\mu $ in Eq.~(\ref{eq2}) as follows
\begin{equation}\label{eq14}
P \beta= \mu^2;  \qquad P=  \int_{0}^{+\infty} x^{-5/2-n} \Phi(x) dx
\end{equation}
The integral term $P$   can be easily calculated numerically or even analytically \citep[cf.][Eqs.~(2.2, 5.7, 7.10)]{GlazmanSrokosz1991} for a particular spectral shape function. For wind wave spectra the result of the integration depends weakly on stage of wave growth, the corresponding property is known as\emph{ spectral shape invariance } \citep{HassEtal73,Hass_ross_muller_sell76}. Taking this property into account one can easily show that terms $\lambda_0,\,\lambda_1$ in Eq.~(\ref{eq10}) and the resulting SSB can be written in the form
\begin{equation}\label{eq15}
  \lambda_0 \sim A(\xi)\mu; \quad \lambda_1 \sim B(\xi)\mu; \quad SSB \sim C(\xi) \mu H_s
\end{equation}
where wave steepness $\mu$ describes general linear dependence of the skewness bias (\ref{eq10}) while functions of pseudo-age $A(\xi),\,B(\xi),\,C(\xi)$ represent  the property of spectral shape invariance. The relatively weak variations of the coefficients  of proportionality $A(\xi),\,B(\xi)$ in Eq.~(\ref{eq13}) { and, hence,  $C(\xi) $ in Eq.~(\ref{eq10})  with} stage of wave development can be quantified using empirical dependencies of wave spectra parameters (shape function $\Phi$) on sea state \citep[e.g.][]{Babanin_Soloviev98a}.

Eq.~(\ref{eq15}) provides a ground for the theoretical assessment of the skewness bias associated with the altimeter-derived wave steepness $\mu$. { As far as  wave steepness $\mu$ dramatically affects the wave breaking and, thus, the EMB, the experimental observation of the proportionality laws given by Eqs.~(\ref{eq15}) can be questionable \citep[cf.][]{GommenBias2003}.}

\section{SSB data of Jason-3 and SARAL/AltiKa within dimensional and dimensionless formulations}
\label{sect3}

This section is related to carry out the SSB analysis with respect to two pairs of parameters: dimensional wave height and wind speed, and dimensionless pseudo-wave age $\xi$ and wave steepness $\mu$. Jason-1,2,3 and SARAL/Altika missions are taken into consideration as testing datasets.  It should be stressed, that the SSB series of the SGDR Level 2 products cannot be considered as `an ultimate truth'. The corresponding parametric models provide the best fit in space of dimensional $H_s$ and $U_{10}$ but, in general case, do not guarantee a  reasonable result  in space of dimensionless pair $\xi,\,\mu$. Moreover,  wind speed $U_{10}$  is taken from empirical models and represents \emph{the most probable value} in contrast to \emph{instantaneously measured } wave height $H_s$ and its gradient $\nabla H_s$. Thus, dimensionless  pair $\xi,\,\mu$ provides rather physical than statistical method of  SSB assessment \citep[see discussion][]{BadulinEtal2018}.

Firstly, we focus on specific properties of wave steepness $\mu$  (Eq.~\ref{eq3} ) as an argument of the SSB parametric model.   The universality of its geographical and statistical distributions and weak dependence on the altimeter band (Jason vs SARAL/AltiKa missions) makes it a reliable and robust predictor for SSB.

The next step of the analysis is finding out the similarity and robustness of SSB distributions themselves as functions of the newly proposed variables.

\subsection{Data description}
%%%%%%
The data used here has been retrieved from Sensor Geophysical Data Records (SGDR, Level 2 product) with 1-Hz sampling  from the AVISO FTP website (\verb"https://www.aviso.altimetry.fr"). The analysis was conducted for the 2018 year: full cycles $70-105$ for Jason-3 and cycles $116-124$ for SARAL/AltiKa mission. The data provided the following  parameters for further analysis:
\begin{enumerate}
\item Ka-band and Ku-band corrected (MLE3 retracking)	significant wave height $H_s$;
\item	sea state bias (SSB) correction in Ku- (Jason-3) and Ka-bands (SARAL/AltiKa);
\item altimeter wind speed;
\item coordinates (longitude, latitude) and time of standard 1-sec altimeter records.
\end{enumerate}
Comparative analysis of the missions was mostly focused on the relevance of the dimensionless variables $\xi,\,\mu$ for the problem of SSB.
Data of two pairs of missions Jason-1--Jason-2 (cycles $257-294$ and $18-55$, year 2009, correspondingly)  and Jason-2--Jason-3 (cycles $18-55$ for 2009 and $70-105$, year 2018, correspondingly) were analyzed in order to exclude (or minimize) the effect of device dependence and to find out the similarity of the SSB distributions.

Following recommendations of \citet{GavrikovEtal2016} the source track data were filtered. Along-track derivative of $H_s$ in Eq.~(\ref{eq3}) and the corresponding wave steepness $\mu$   was estimated as simple differences between two consecutive track points.

All data passed quality control procedure. After being filtered using the corresponding fields, e.g. \verb"ocean" for variable  \verb"surface_type", \- \verb"good" for \verb"qual_alt_1hz_swh_ku_mle3" etc., noises were identified and excluded. Additionally, the homogeneity quality control check  eliminated pseudo-age parameter over a $95$ percentile. Hereafter, the computations were done for $0.5 {\rm m} <H_s < 8$m, $|SSB|<0.5$m,  $1{\rm m/s} < U_{10} < 20{\rm m/s}$. Only ice-free data for latitudes below $60^\circ$ (N, S) were taken for the analysis.

Here we do not present analysis of the measurement errors of $H_s$ and accuracy of estimates of $U_{10}$ referring to generally accepted values $0.4$ m and $1.5$ m/s, correspondingly \citep[e.g.][]{RussianSeas2011,HandbookJason32016}.

\subsection[General characteristics]{General characteristics of altimeter-derived parameters}
%%%%%%%%%%%%%%%%%%
\begin{figure}
\centering
\includegraphics[scale=0.75]{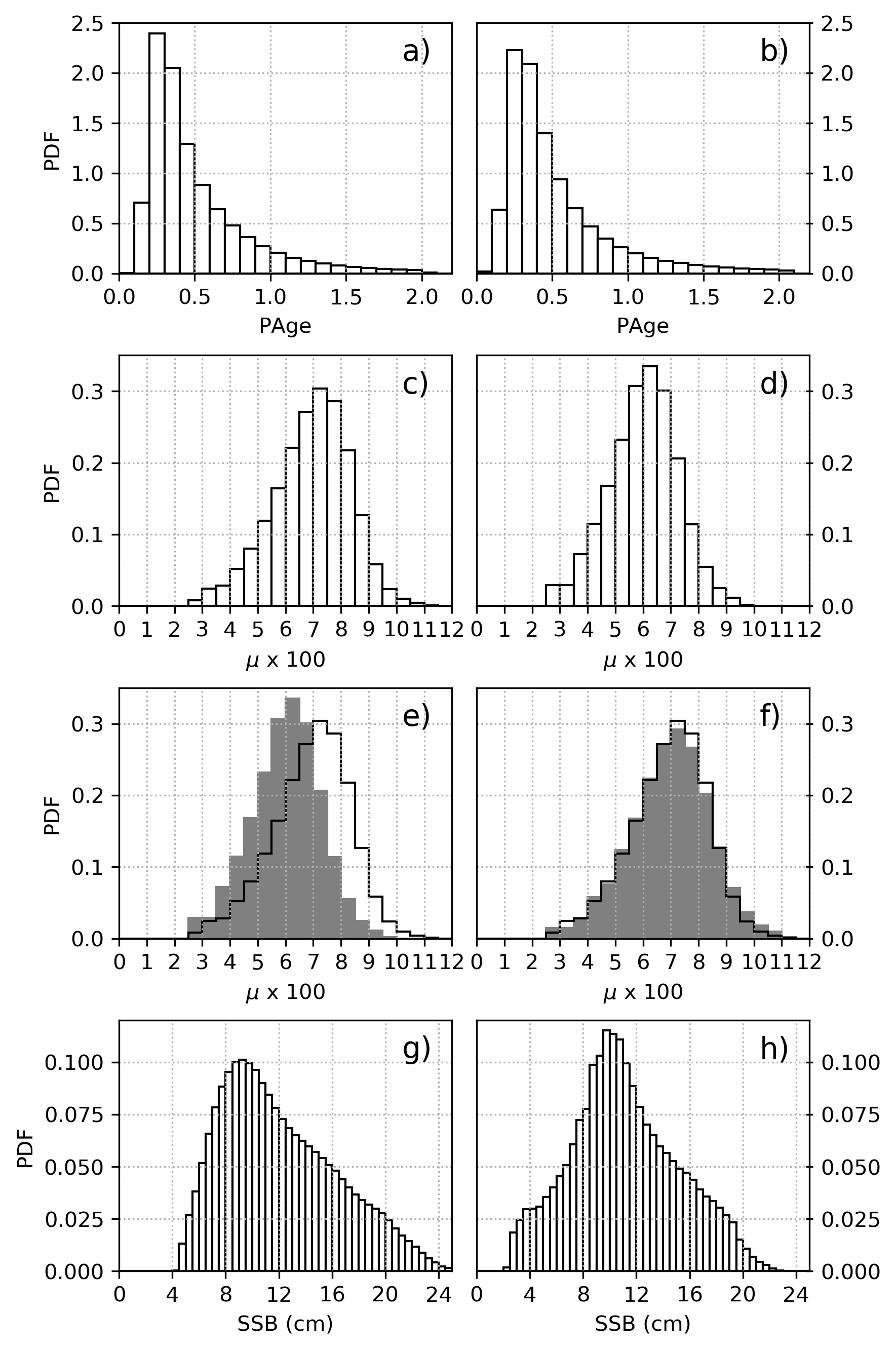}\\
 \caption{Probability density functions of altimeter-derived dimensionless values for Jason-3 (left column) and SARAL/AltiKa (right column) for the  year 2018. a,b) -- pseudo-age $\xi=gH_s/U_{10}^2$ ;  c,d) -- wave steepness $\mu$; e) -- comparison of PDFs of two missions: unshaded curve -- Jason-3, shaded -- SARAL/AltiKa; h) -- the same as h) with the re-scaled $\mu$ of SARAL/AltiKa ($\mu_{new}=1.159 \mu$); g,h) -- normalized sea state bias $SSB/H_s$. }
\label{fig1}
\end{figure}

As a preliminary step of the analysis, the normalized probability density functions (PDF) have been calculated for key measurable and derived parameters. Fig.\ref{fig1} shows results for the pair of missions Jason-3 -- SARAL/AltiKa, year 2019.

Significant wave heights of both missions show  similar  smooth distributions  with a well-pronounced maximum of about  $2$ m  (not shown here). Wind speed PDFs (not shown) demonstrate some discrepancies in shape, especially for $0.5$ m/s bin size, which is, however,  less than the accuracy of wind speed in altimeter measurements \citep{RussianSeas2011,BonnefondCoastal2011}. The  most probable values of both missions remain close and fall within $6-8$ m/s. Dimensionless pseudo-age $\xi$ (Fig.\ref{fig1}{\it a,b}) also demonstrates an apparent agreement for  both missions.

\begin{figure}
\centering
\includegraphics[scale=0.35]{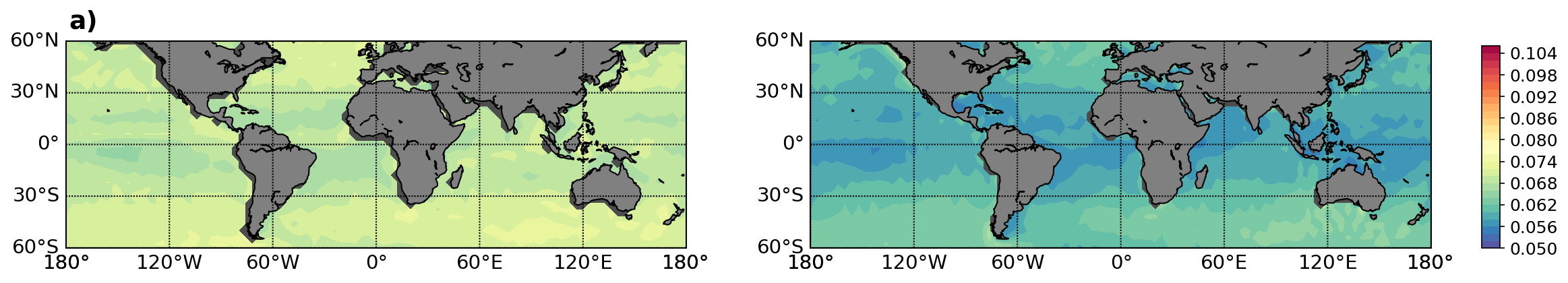}\\
\includegraphics[scale=0.35]{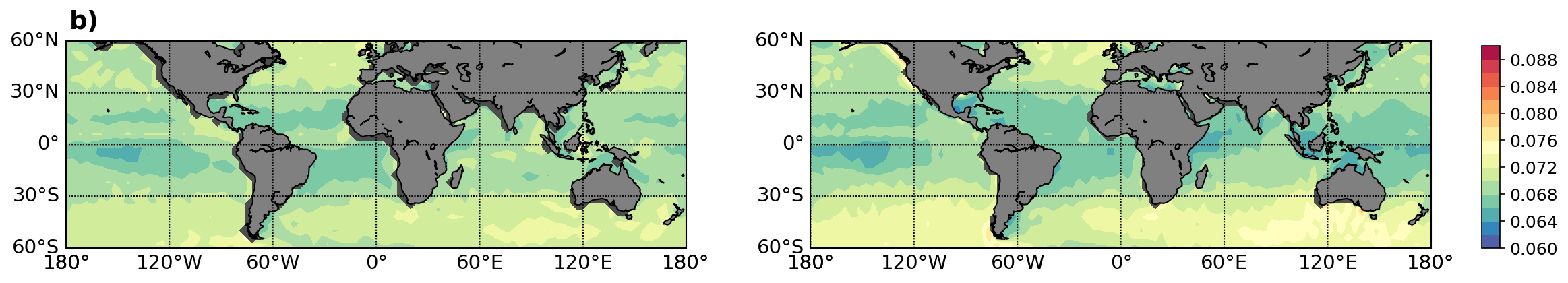}\\
\includegraphics[scale=0.35]{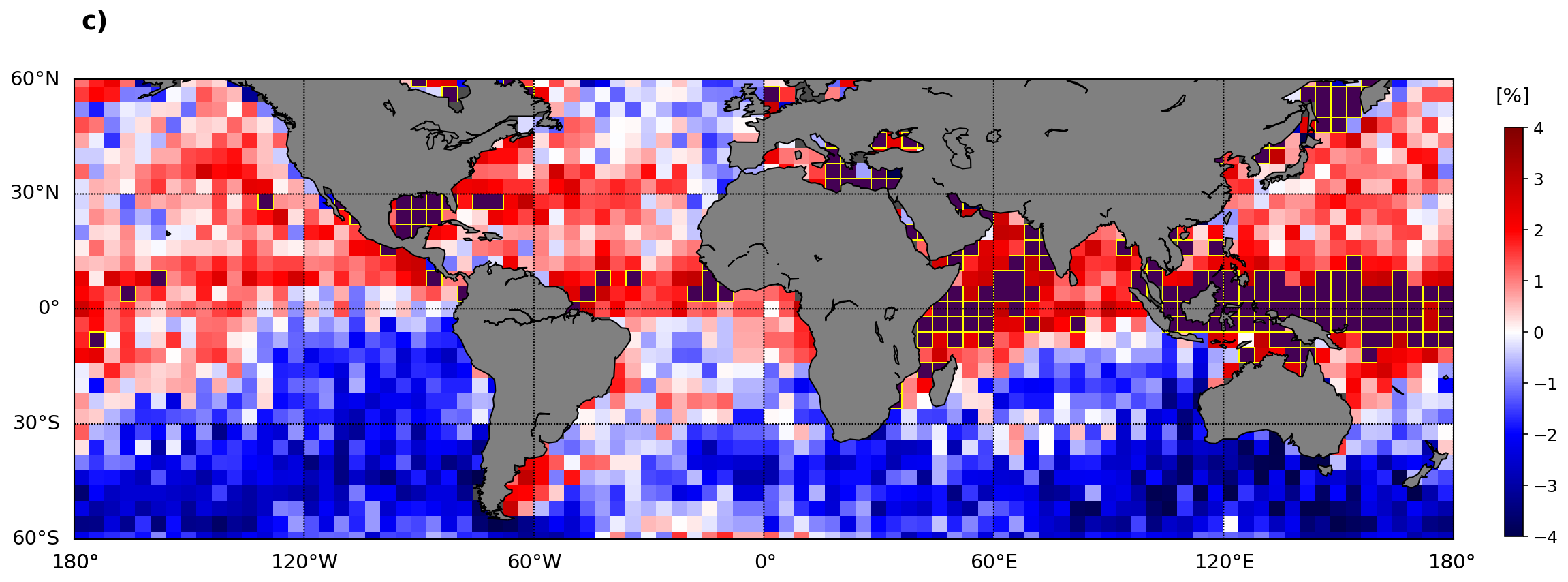}
 \caption{a) -- global distribution of mean over $4^\circ \times 4^\circ$ coordinate boxes wave steepness $\mu$  Jason-3 (left column) and SARAL/AltiKa (top panels). b) -- the same for SARAL/AltiKa wave steepness re-scaled by factor $1.159$. c) -- normalized difference of wave steepness of Jason-3 and re-scaled SARAL/AltiKa defined by Eq.~(\ref{eq16}). The outliers exceeding $4\%$ are shown in violet.}
\label{fig2}
\end{figure}

PDFs of wave steepness $\mu$ estimated by Eq.~(\ref{eq3}) (Fig.\ref{fig1}{\it e,f}) differ in the magnitude and peak positions but still look similar in shape. This similarity can be justified with a simple one-parametric transformation. Multiplication of the SARAL/AltiKa estimates of $\mu$ by a factor $1.159$ makes both distributions identical in the whole range of $\mu$ with the perfectly matched peak positions  (Fig.\ref{fig1}{\it e,f}).   Such a remarkable transformation  can be explained by the geometry of satellite tracks \citep[see discussion in ][]{BadulinEtal2018}. The near-pole orbit of SARAL/AltiKa (inclination $98.54^\circ$) makes estimates of $\mu$ more sensitive to predominantly zonal directions of wind and waves. Thus, SARAL/AltiKa tends to underestimate $\mu$ by approximately $15\%$  of magnitudes as compared to Jason-3 with the  orbit inclination $66^\circ$.
The  similarity of wave steepness  PDFs  of two missions is  proved by global distributions of wave steepness in Fig.\ref{fig2}. Being originally close in spatial patterns (Fig.\ref{fig2}{\it a}) Jason-3 and Saral/AltiKa demonstrate an impressive quantitative agreement  after re-scaling procedure in Fig.\ref{fig2}{\it b}.  Averaged over $4^\circ \times 4^\circ$ boxes $\mu $ varies in rather narrow range (Figs.\ref{fig2}{\it a,b}), extreme values differ by less than $20\%$ of maximal magnitudes for both missions. Recently, this  general property   has been treated as a universality of `wave steepness climate' by \citet{BadulinEtal2018}.

Fig.\ref{fig2}{\it c} presents normalized deviation of estimates of $\mu$ for Jason-3 ($\mu_{J3}$) and corrected SARAL/AltiKa ($\mu_{SA}$). This deviation defined as
 \begin{equation}\label{eq16}
   \Delta \mu=100\times \frac{\mu_{J3}-\mu_{SA}}{\mu_{SA}}
\end{equation}
 does not exceed $4\%$ of local magnitudes in $95\%$ of coordinate boxes in Fig.~(\ref{eq2}). Note, that the deviation is mostly positive in the Northern Hemisphere (wave steepness of Jason-3 is higher than one of SARAL/AltiKa) and negative in the Southern Hemisphere. Additional efforts are needed to make clear the origin of these differences.

 The consistency of both missions in terms of PDFs of wave height $H_s$, wind speed $U_{10}$, pseudo-age $\xi$ and wave steepness $\mu$ appears in visible contrast with the SSB estimates (fig.\ref{fig1}{\it g,h}). Jason-3 with its long-time history of follow-on missions (since launching TOPEX/Poseidon  in 1992) provides a smooth PDF. Oppositely, the relatively short record of the SARAL/AltiKa shows a rather sharp distribution with shifted probability maximum and spurious tails.

Similar  statistical distributions for missions Jason-1,2 (not shown here)  are in  perfect agreement. In fact, this is because that follow-on missions of Jason instruments have the same thorough calibration and data retracking methods with close characteristics.

The above simple analysis of global distributions provides a ground for discussion of similarity and deficiency of SSB models in terms of conventional dimensional ($H_s,\,U_{10}$) and dimensionless ($\xi,\mu$) values.

\subsection{Conventional approach for SSB: dimensional arguments $H_s$ and $U_{10}$ }
%%%%%%%%%%%%%
\begin{figure}
\centering
\includegraphics[scale=0.225]{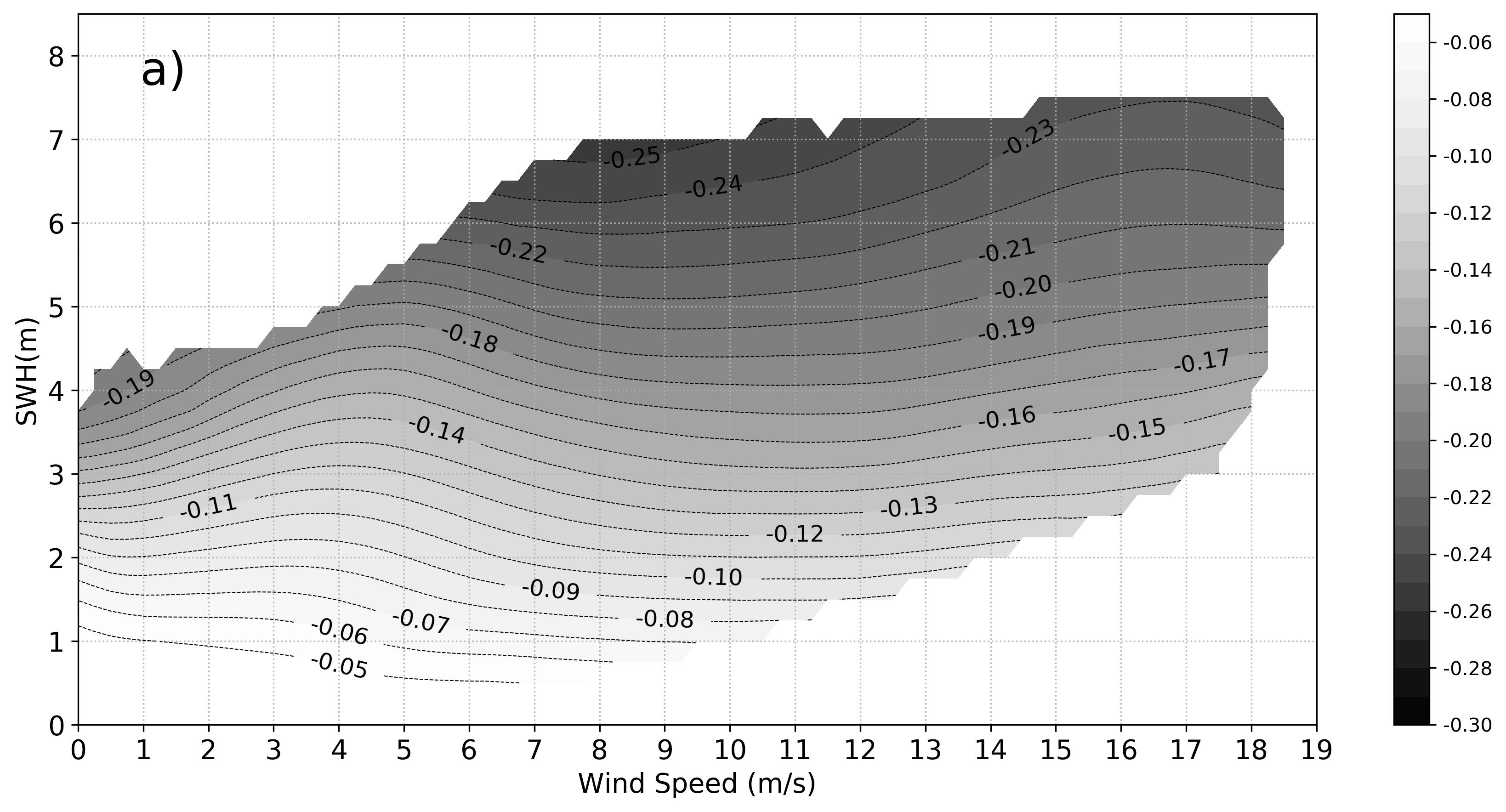}
\includegraphics[scale=0.225]{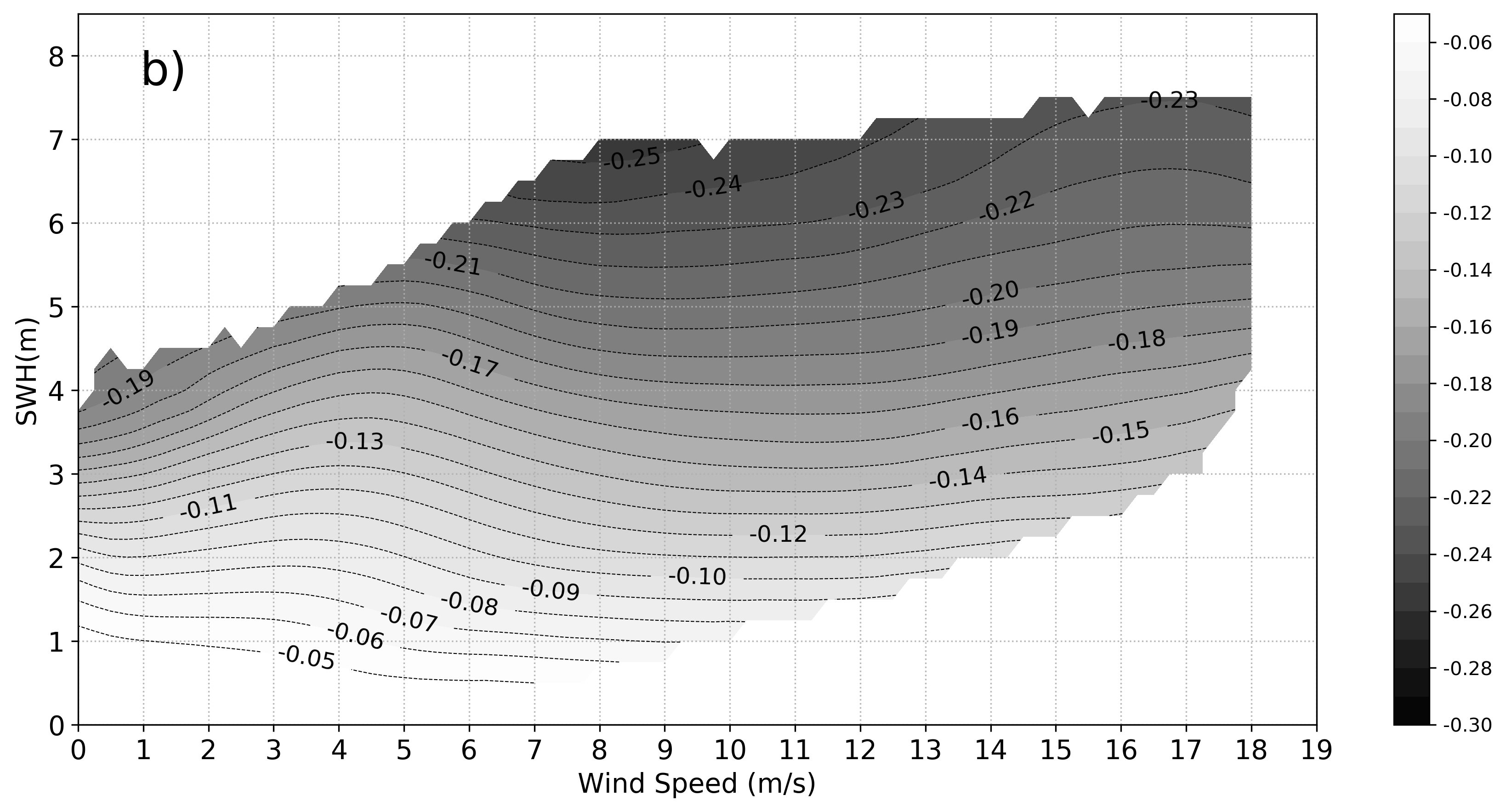}\\
\includegraphics[scale=0.225]{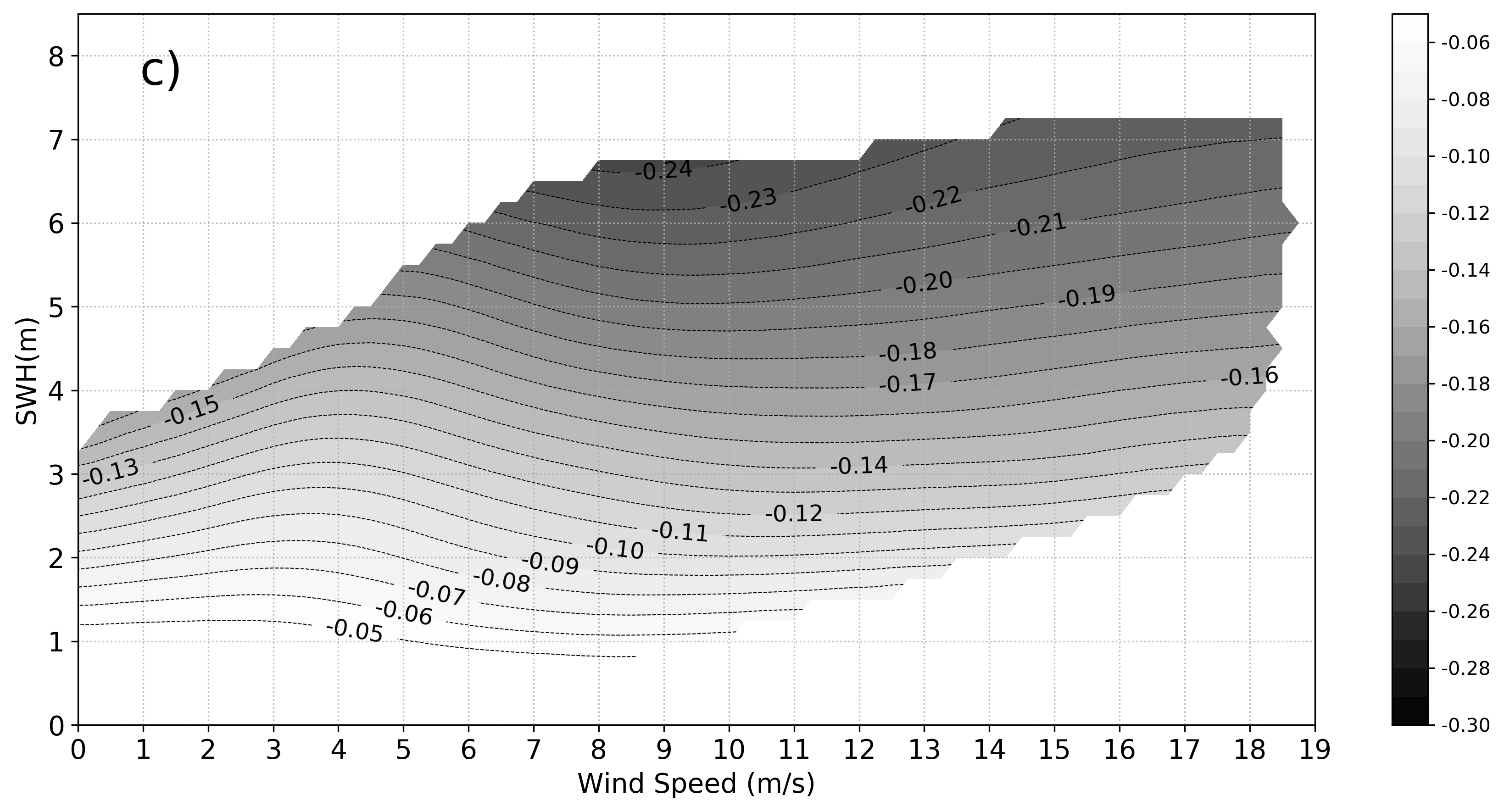}
\includegraphics[scale=0.225]{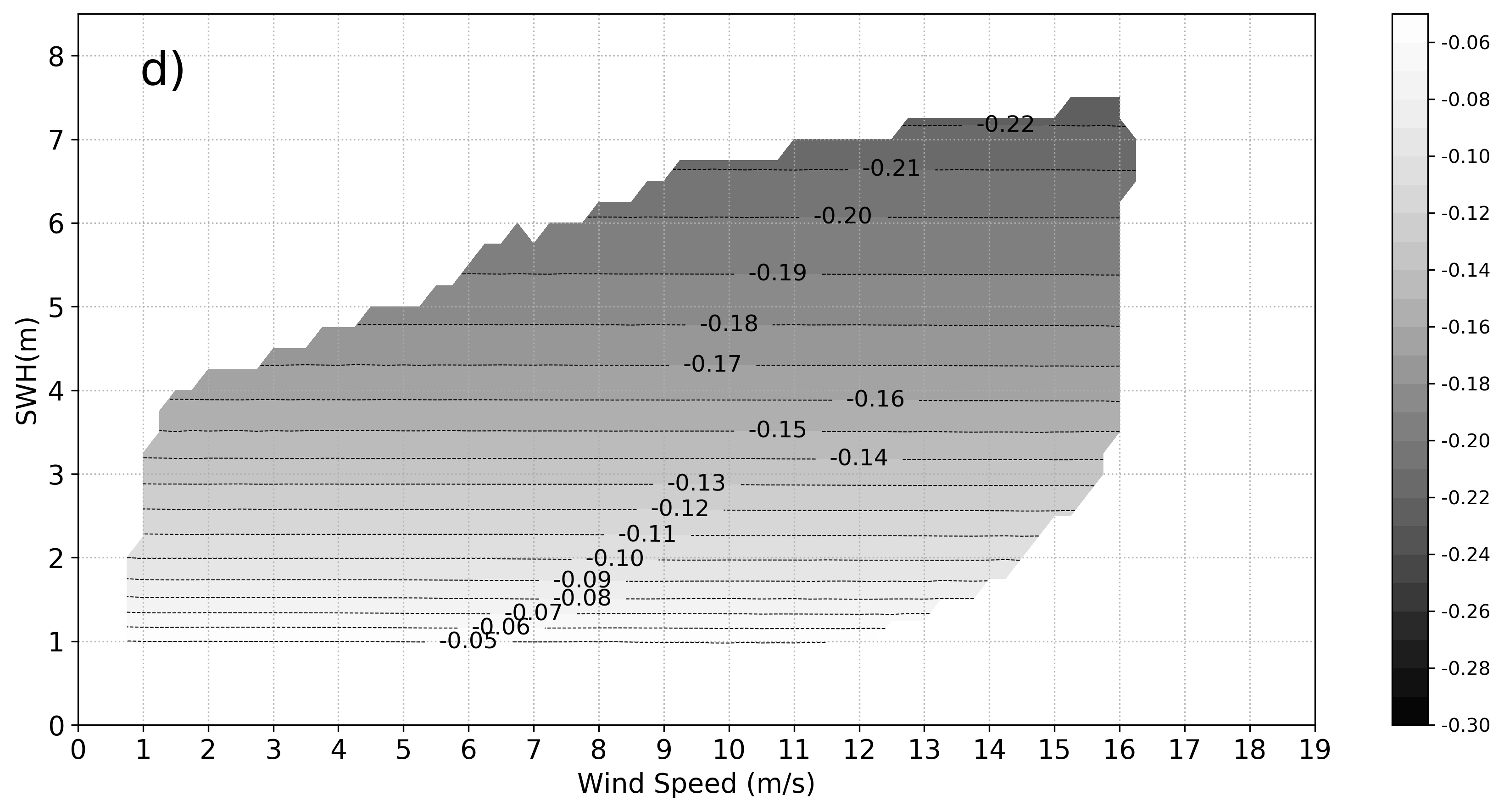}\\
\includegraphics[scale=0.225]{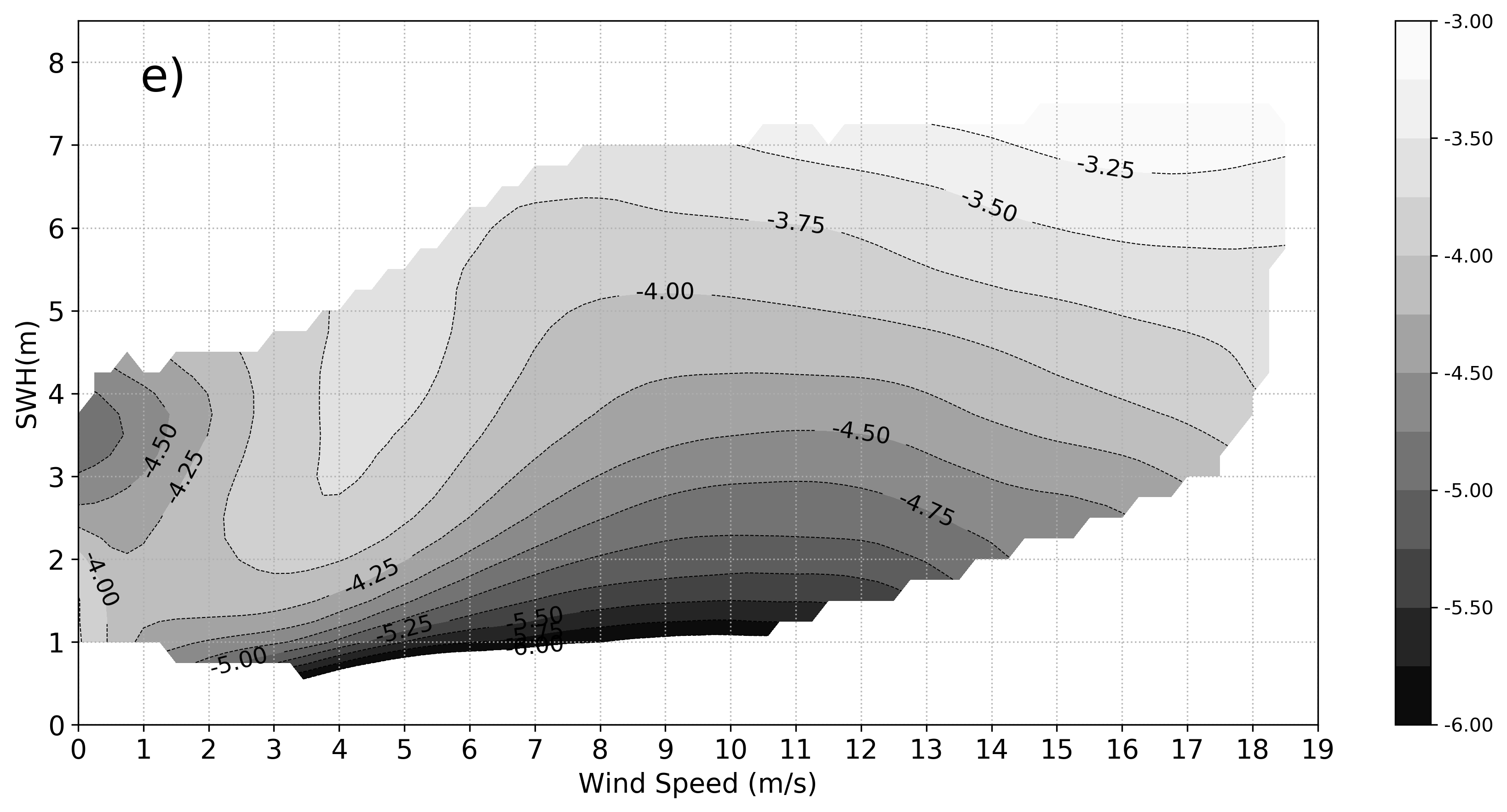}
\includegraphics[scale=0.225]{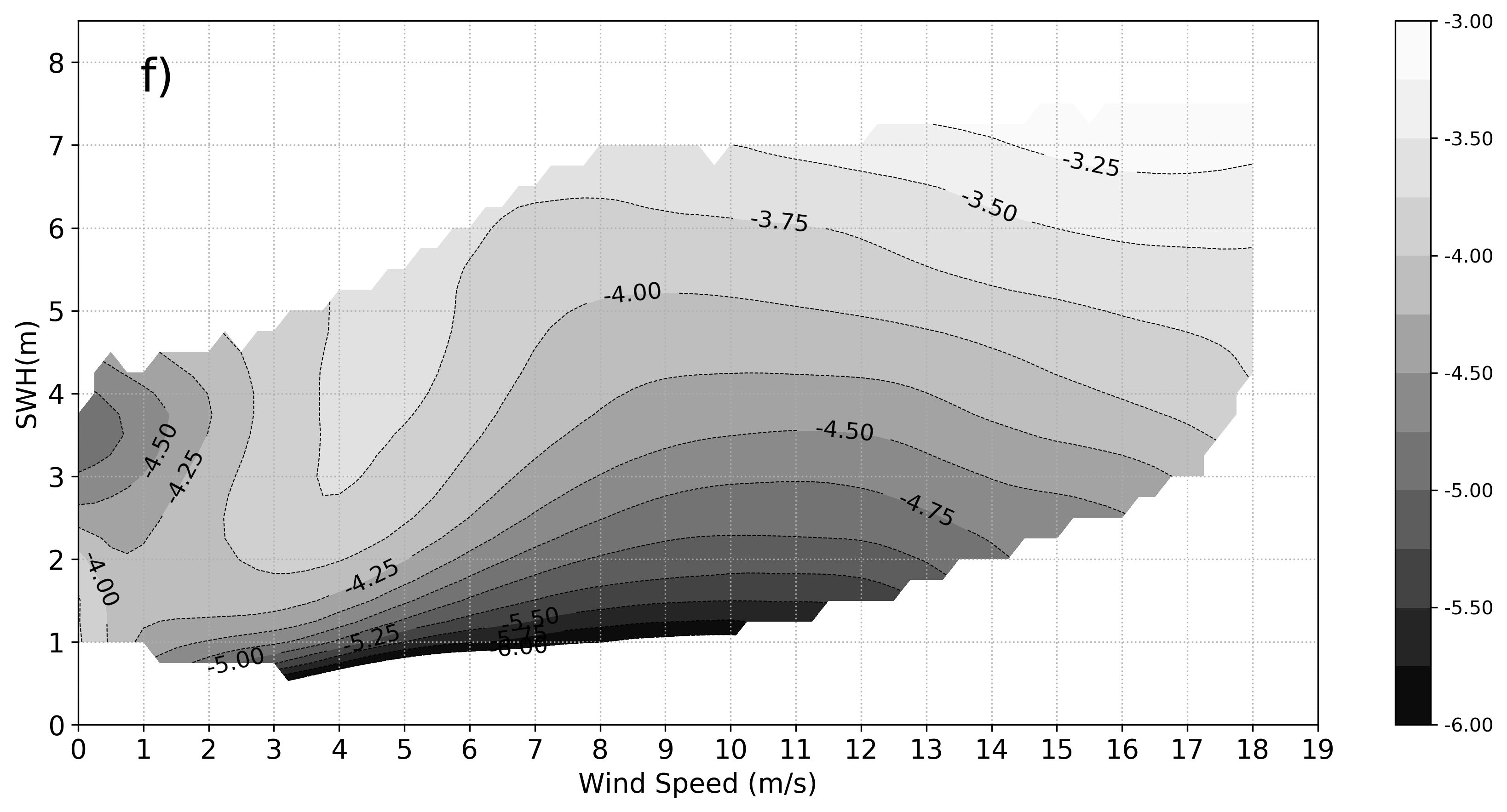}\\
\includegraphics[scale=0.225]{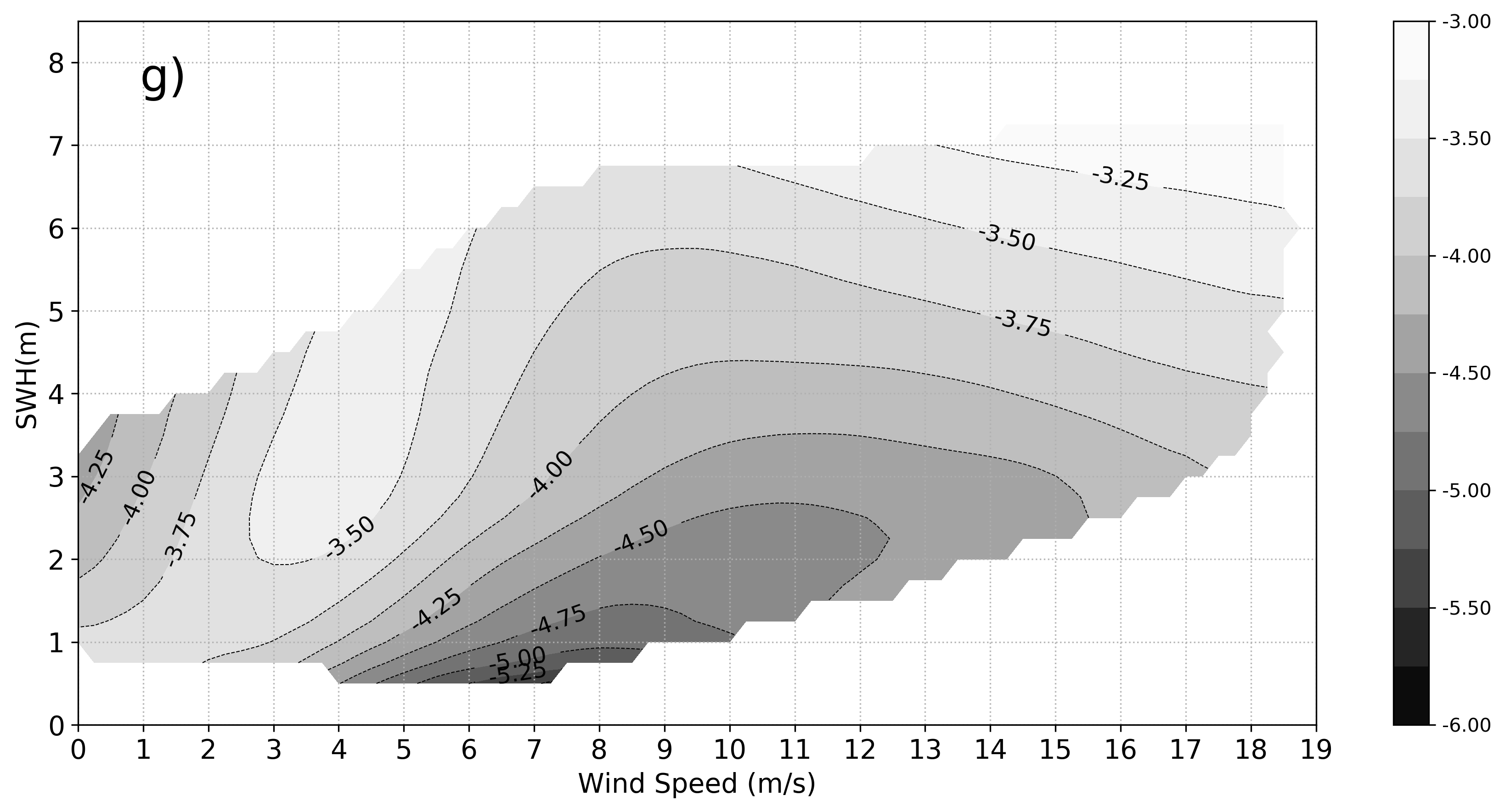}
\includegraphics[scale=0.225]{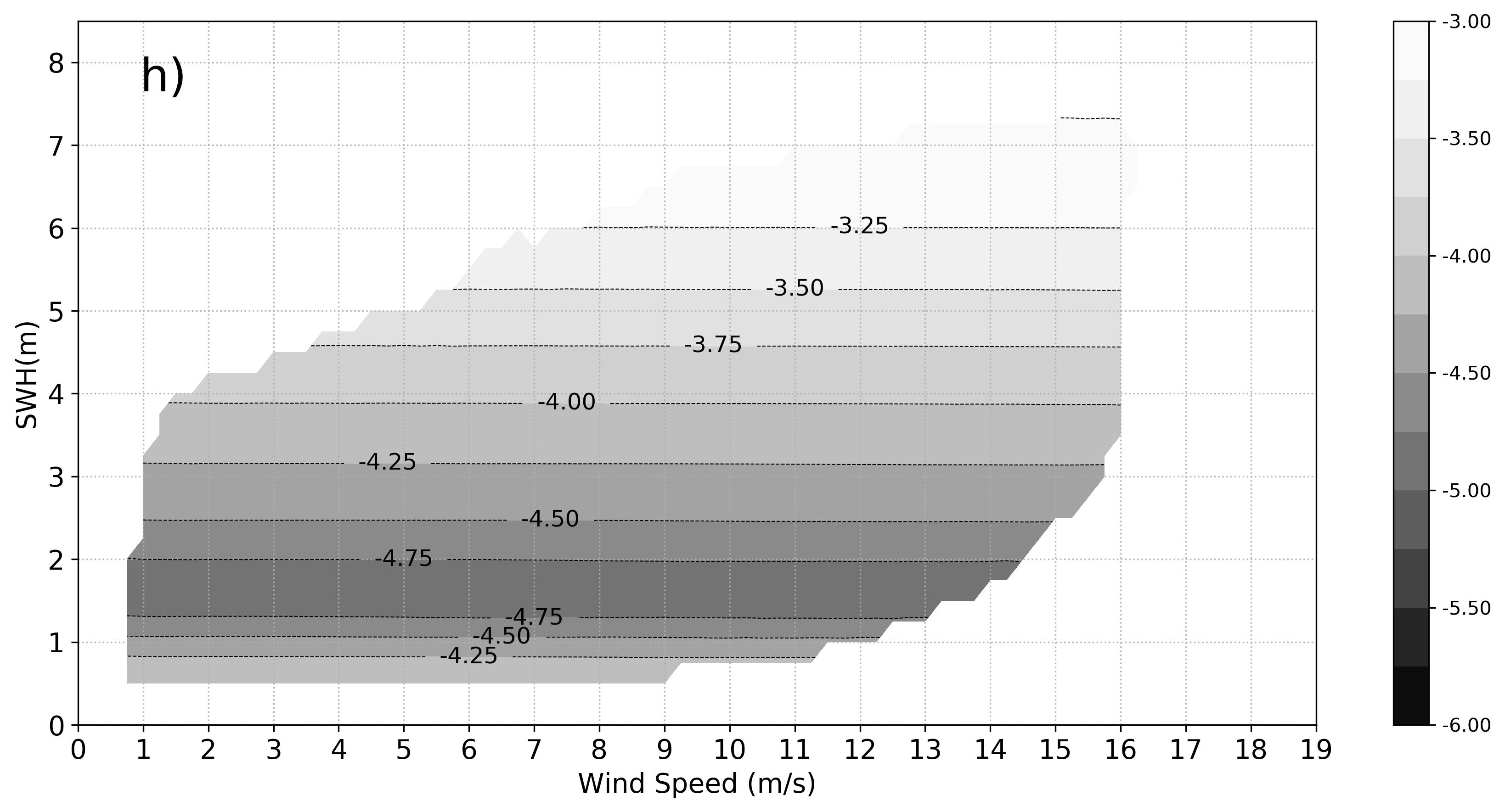}
\caption{a,b,c,d) -- isolines for the global  $SSB$  (in meters)  obtained from bin-averaging into boxes of width ($0.25$ m/s, $0.25$ m) over the ($H_s,U_{10}$) domain. a)  -- Jason-3, cycles $70-105$, b) -- Jason-2, cycles $18-55$; c) -- Jason-1, cycles $257-294$; d) -- SARAL/AltiKa, cycles $116-124$   \citep[cf.][Fig.1]{Vandenmark2002};
e,f,g,h) --  isolines for normalized value $SSB/H_s$ (in percents) over the ($H_s,U_{10}$) domain for the same data series.}
\label{fig3}
\end{figure}
Efficiency of the SSB mapping as a function of dimensional arguments $H_s,\,U_{10}$ has shown its practical value \citep[e.g.][and refs. therein]{PiresEtal2018} when comparing with non-parametric crossover-based models. Fig.\ref{fig3}{\it a,c,d} illustrates a transparent consistency of $SSB(H_s,U_{10})$ distribution for Jason-1,2,3 and  one for TOPEX/Poseidon \citep[cf.][Fig.1]{Vandenmark2002}. Bin sizes $0.25$m/s in $U_{10}$ and $0.25$ m in $H_s$ were taken the same as in the cited paper. All the patterns demonstrate a gradual increase of $SSB$ with the increasing $H_s$ and wind speed. Local peculiarities at high values of wind speed and wave height in \citet[][Fig.1]{Vandenmark2002} are smoothed in the Jasons' patterns. The found correspondence of Jason-1,2,3  as  follow-on missions of the TOPEX/Poseidon seems  natural because of a long duration of missions and great amount of data underlying the estimates of SSB.

Similar patterns for dimensionless $SSB/H_s$ are identical for Jason-2 and 3 \citep[cf. Fig.~\ref{fig3}{\it e,h} and sects. 5.6.3 in][]{J2Manual,HandbookJason32016}. However, difference  of the distribution of $SSB/H_s$ for Jason-1 caused by different retracking methods (MLE3 retracking for Jason-2,3) become more pronounced.

The SARAL/AltiKa mission (Fig.\ref{fig3}{\it b,d}) with its relatively short duration  of measurements portrays a different pattern when the SSB does not depend on wind speed $U_{10}$. Physical treatment of  both  Jason and SARAL/AltiKa distributions in terms of dimensional $H_s,\,U_{10}$ is quite simple: higher waves as well as stronger winds provide higher SSB. Collecting more data could  improve the situation.
\begin{figure}
\centering
\includegraphics[scale=0.35]{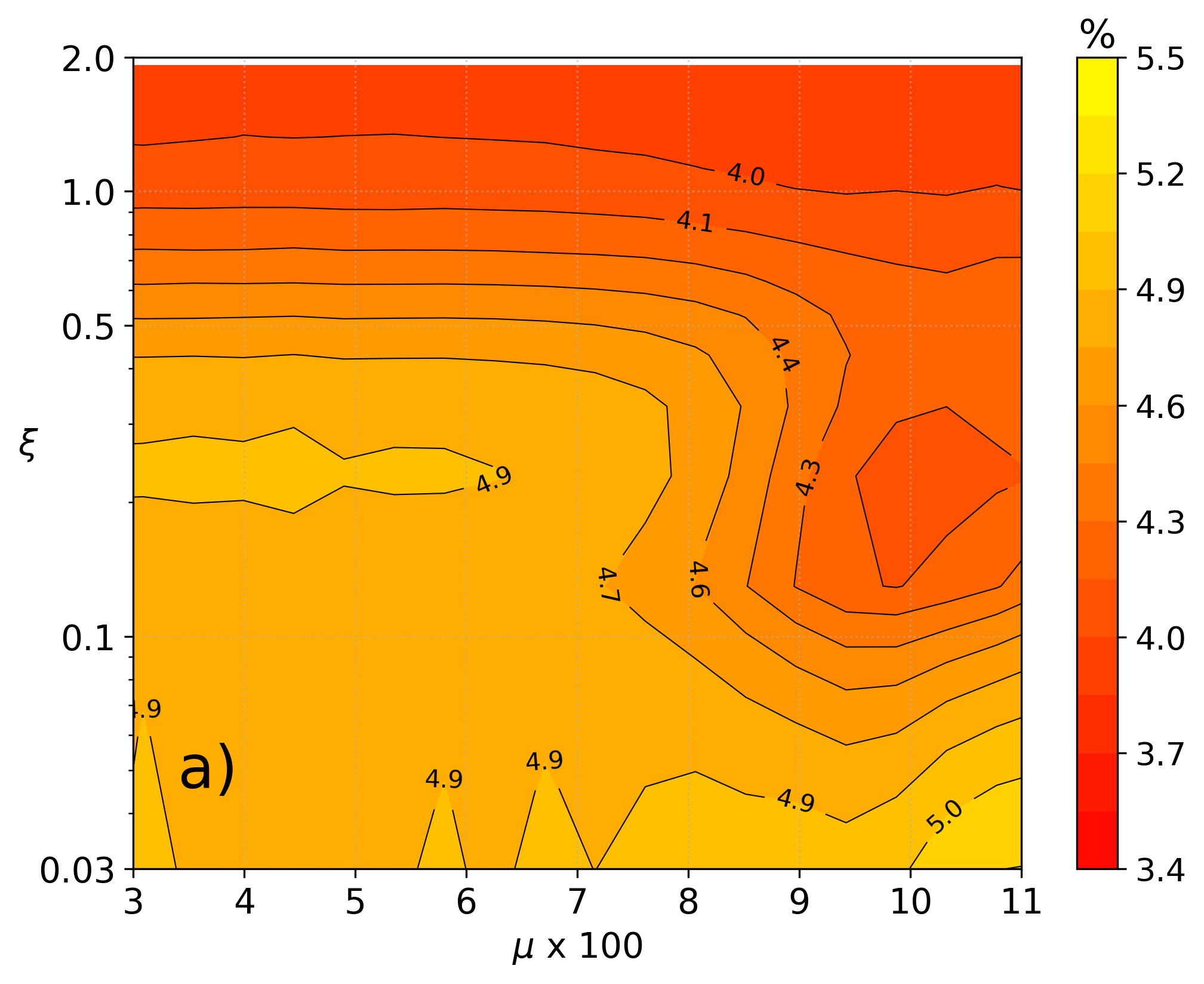}
\includegraphics[scale=0.35]{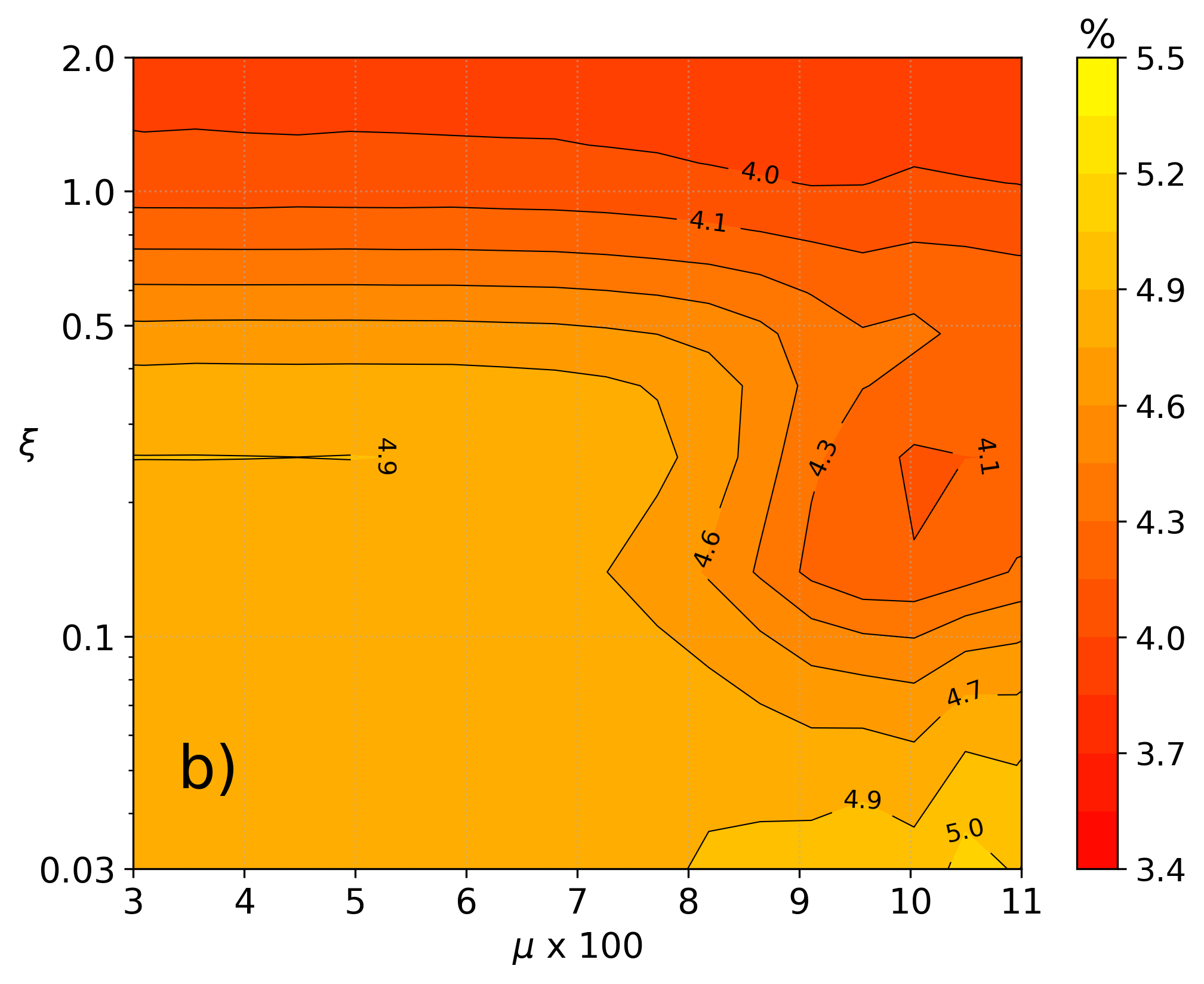}\\
\includegraphics[scale=0.35]{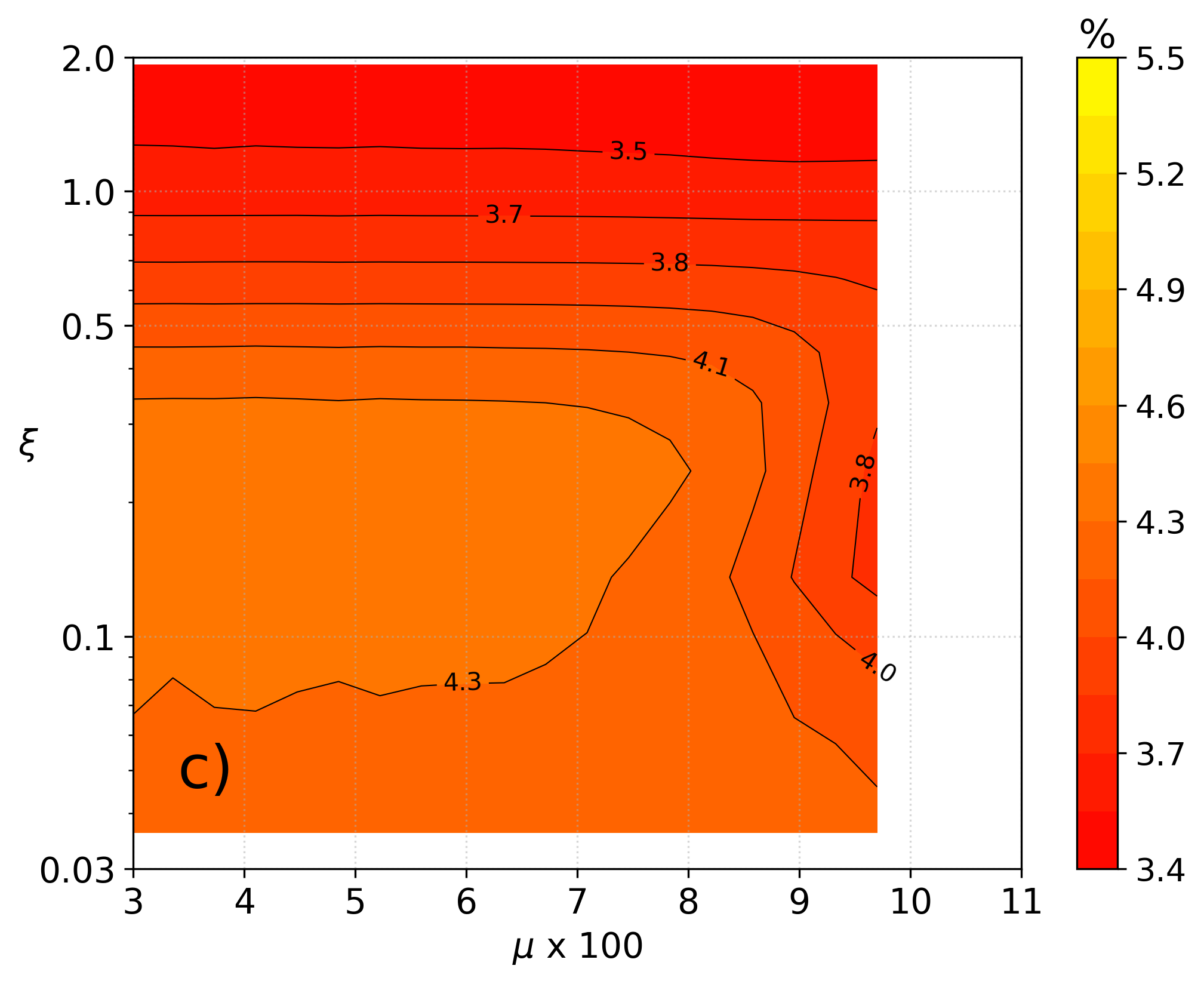}
\includegraphics[scale=0.35]{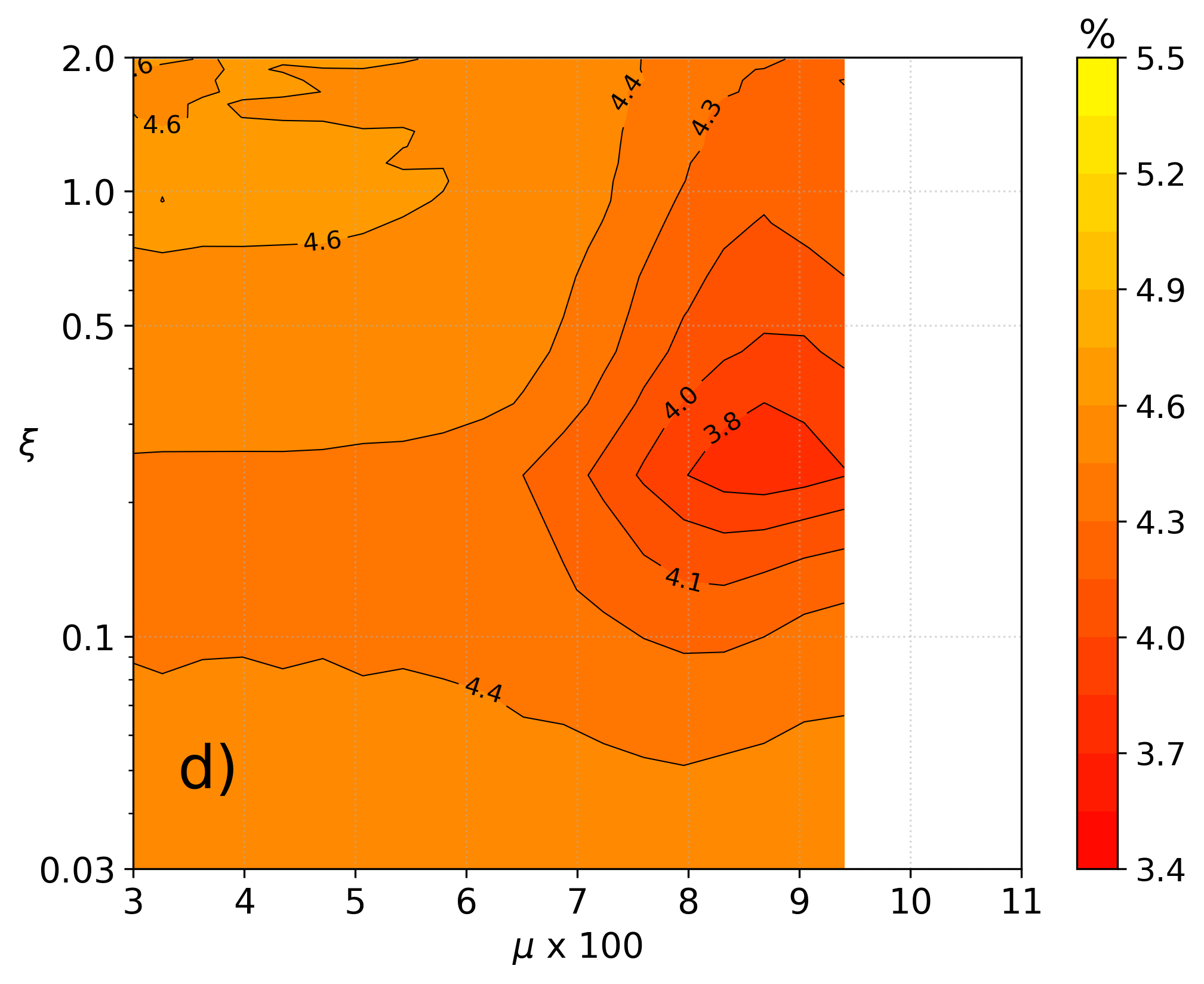}
\caption{Isolines for the global  normalized value $SSB/H_s$ (in percents) obtained from bin-averaging into boxes of width $\Delta \xi/\xi =1.1$, $\Delta \mu =0.002$. a)  -- Jason-3, cycles $70-105$, b) -- Jason-2, cycles $18-55$; c) -- Jason-1, cycles $257-294$; d) -- SARAL/AltiKa, cycles $116-124$   }
 \label{fig4}
\end{figure}

\begin{figure}
\centering
\includegraphics[scale=0.35]{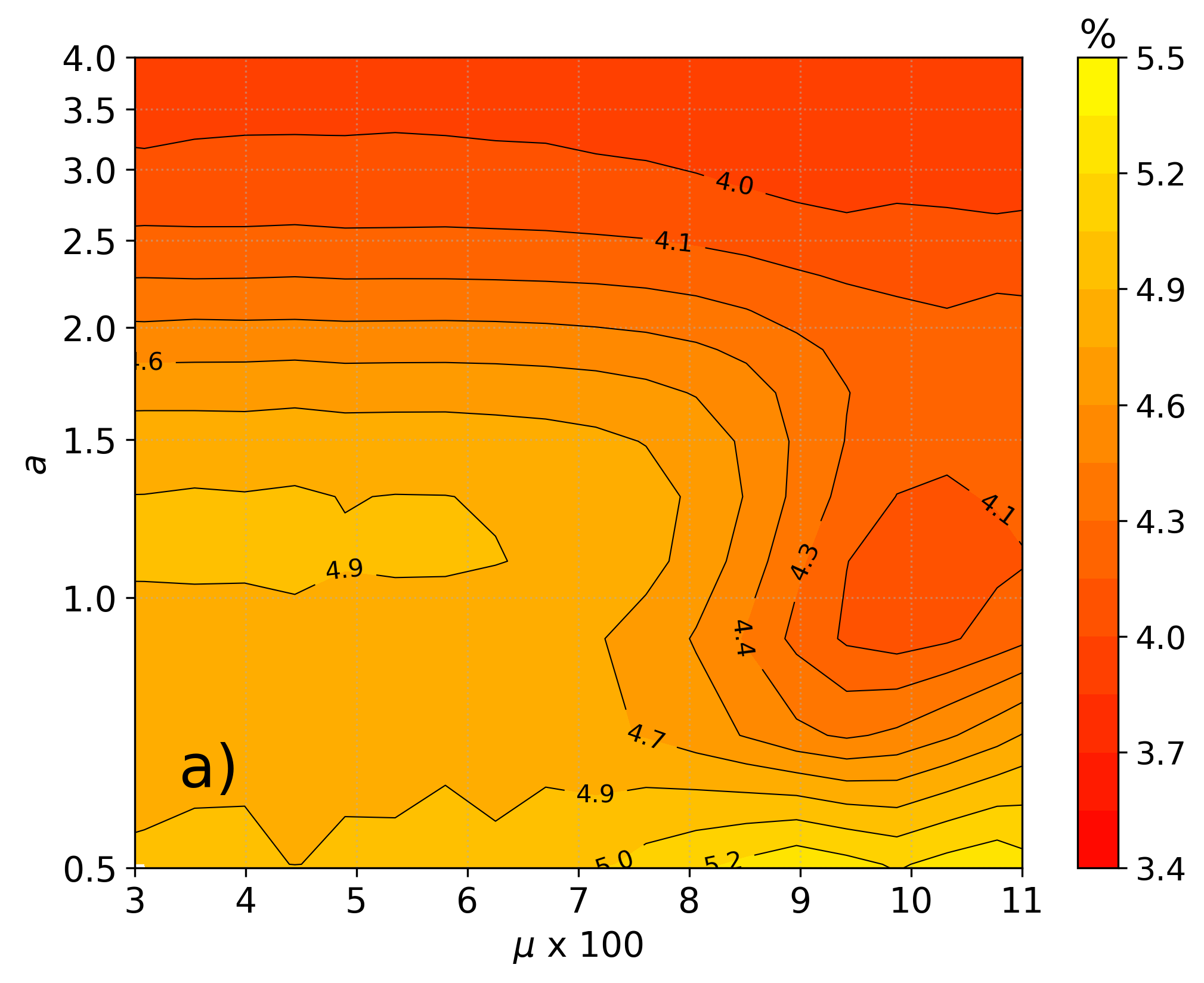}
\includegraphics[scale=0.35]{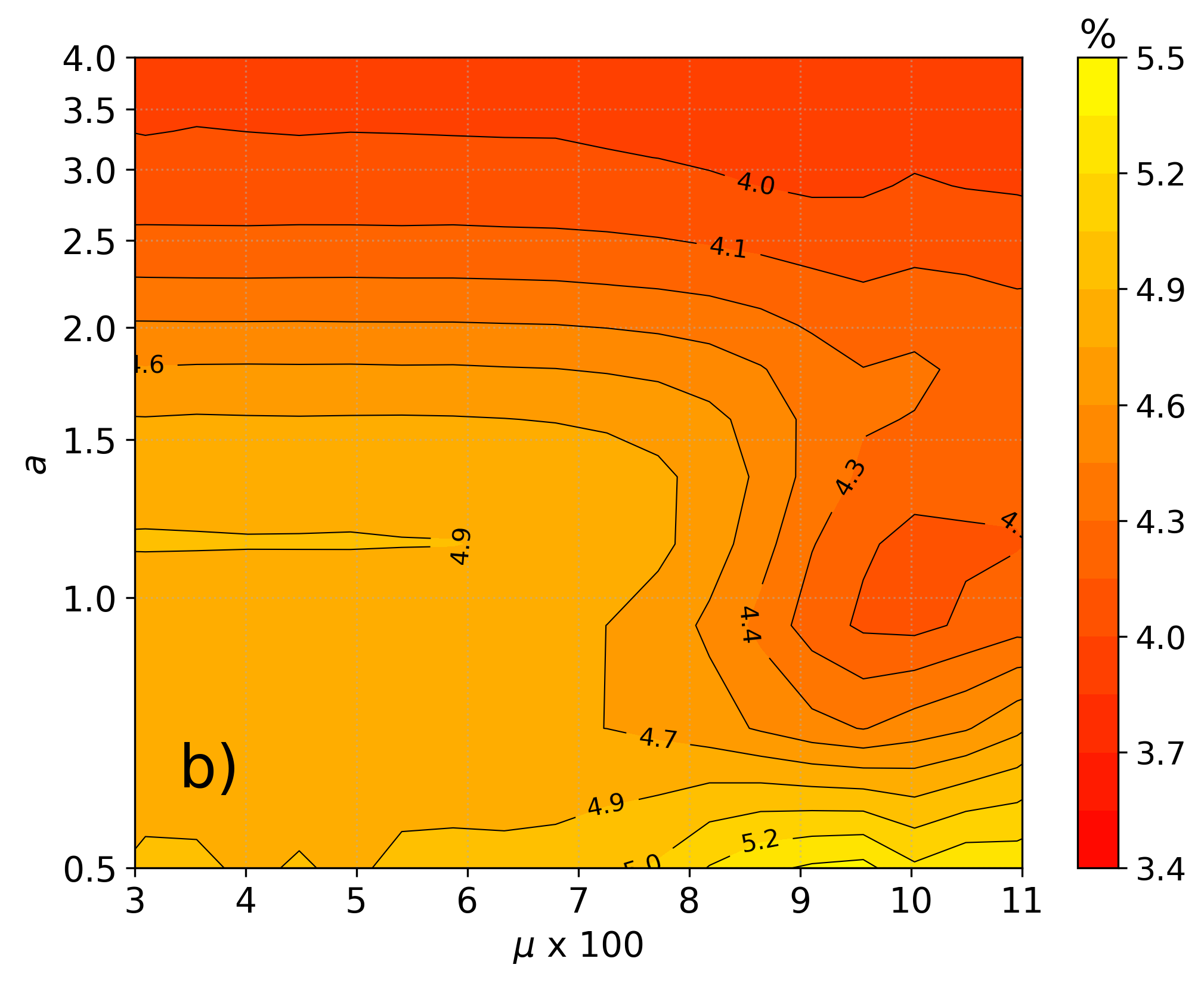}
\includegraphics[scale=0.35]{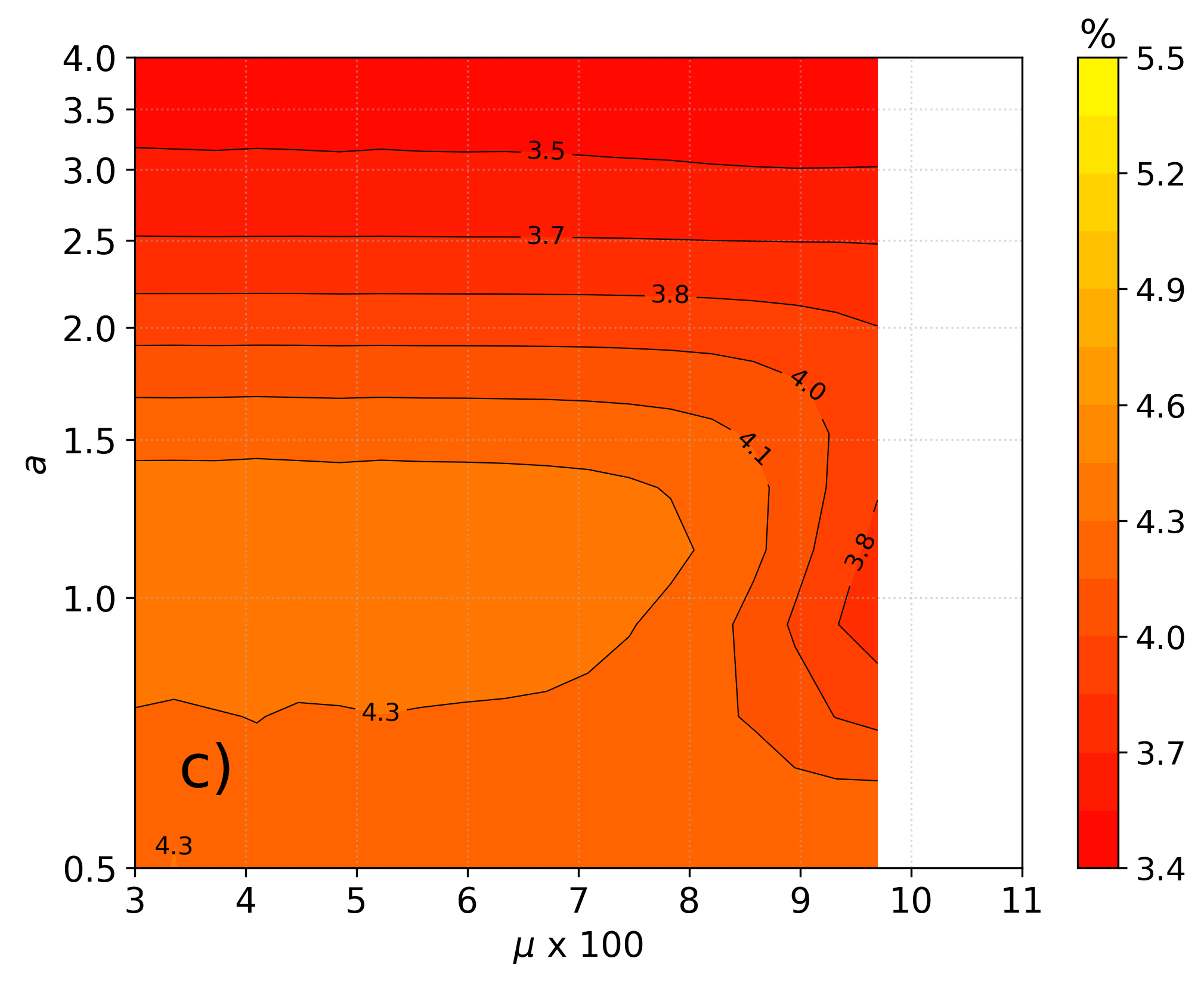}
\includegraphics[scale=0.35]{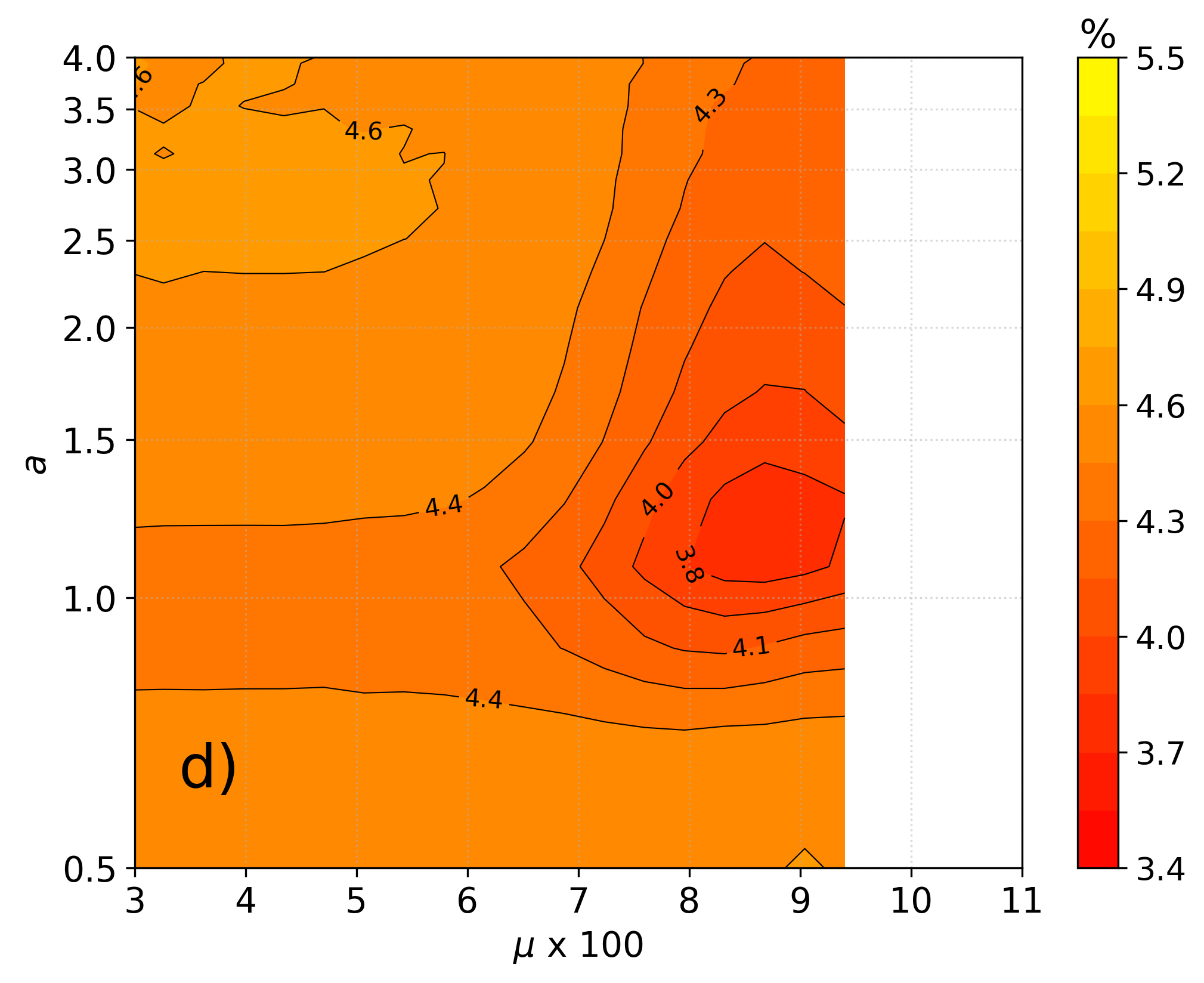}
\caption{Same as previous for wave age $a=gT_p/(2\pi U_{10})$}
\label{fig5}
\end{figure}

\subsection[SSB within dimensionless variables]{SSB within dimensionless  $\xi,\,\mu$: a step towards a physical analysis}
%%%%%%%%%%%%%%%%%%%%%%%%%
As it was mentioned above,  re-casting SSB onto  $(\xi,\,\mu)$ does not guarantee the best data fit in this new argument space. But the analysis of resulting SSB distributions for various data sets provides the basis for further development of the proposed approach.

Fig.\ref{fig4} presents bin-averaged normalized values $SSB/H_s$. Fig.\ref{fig4} for Jasons  and SARAL/AltiKa (Fig.\ref{fig4}{\it d} year 2018, $\mu$ is corrected by factor $1.159$, see sect.~3.2) show different patterns as expected for altimeters operating in different bands. In  turns, all the Ku-band Jason's missions in Fig.\ref{fig4}{\it a,b,c} (Jason-1,  Jason-2, year 2009 and Jason-3, year 2018) show definite  similarity in shapes of SSB patterns (palettes are different in Fig.~\ref{fig4}). Moreover, Jason-2 and Jason-3 demonstrate perfect agreement both in shapes and magnitudes. Different years  (2009 and 2018) make such an agreement to be a strong argument for our approach: various data samples show robustness of distributions of $SSB(\xi,\mu)$.

The partial success with re-mapping all the Jason's missions ($1-3$) can be referred to the effect of re-tracking (TB -- tracking bias): MLE3 retracking data has been applied for Jason-2,3 missions only.

Similar maps for the pair $a,\,\mu$ (wave age -- wave steepness) of all four missions are presented in Fig.\ref{fig5}. While the wave age $a$ is not measured directly an empirical parameterization $a(\xi)$ has been used following \citet{KahmaCalcoen1994} \citep[see also][]{ZBGP2019}. This re-mapping allows for relating SSB with stages of wave growth described in sect.~2.2. All the missions show extrema in classic transition from wind sea to swell at $a\approx 1$.
At $\mu \gtrsim 0.07$ and wave ages $a \gtrsim 0.7$ the dimensionless $SSB/H_s$ shows a breakup for all missions. This domain of ($\xi,\,\mu$)-plane corresponds to `the tongue' of normalized SSB in Fig.\ref{fig3}{\it e,f,g} at relatively low wind speed $U_{10}$ and, hence,  might  indicate  issues in  wind measurements  by altimeters.

The dispersion of normalized SSB (not shown)  is large enough and, in some cases, even comparable with the range of variation of the $SSB/H_s$ itself. The low confidence level accumulates not only the random errors  of $\xi$ and $\mu$ but the uncertainties of ($H_s,\,U_{10}$)-based parametric models.   Complete retracking  of SSB in terms of  the new arguments $\xi,\,\mu$ would be helpful  to resolve this issue.

All Figs.~\ref{fig4},\ref{fig5} do not contain any domains of linear dependence of the normalized value $SSB/H_s$ on wave steepness (see Eq.~\ref{eq15}).  In fact, the model of sect.~2.3 describes just a partial effect of wave dynamics on SSB: skewness bias (SB) due to Stokes-like asymmetry of water wave profiles. The analysis shows that this effect is probably compensated by other effects that also dependent on wave steepness, first of all, by electromagnetic bias (EMB).

\section[Summary]{Conclusions and Discussion}
\label{sect4}
The following list of conclusions and remarks overviews the results of the paper:
\begin{enumerate}
  \item The basic idea of this work was to explore a physical relevance of unconventional parameterization of sea state bias (SSB). Two dimensionless arguments, wave pseudo-age $\xi$ defined by Eq.(\ref{eq4}) and the wave slope $\mu$ (Eq.\ref{eq5}) derived from along-track derivatives, were introduced within the consistent similarity analysis. The new approach has not evolved into alternative model of SSB so far. Such a project requires a complete retracking  a wealth of altimeter data in terms of $\xi$ and $\mu$ instead of the conventional dimensional parameters $H_s$ and $U_{10}$.

  \item One comparative advantage is shown with similarity of wave steepness distributions for Jason missions and SARAL/AltiKa operating at different bands. It should be emphasized that $\mu$ captures instantaneous sea state in contrast to wind speed $U_{10}$ based on the parametric models providing the most probable values.

  \item A potential prospects of new approach were examined on readily available SSB records of recent altimetry missions Jason-1,2,3 and SARAL/AltiKa. The conventional parametric ($H_s,\,U_{10}$)-based models are not obliged to generate any consistent patterns in ($\xi,\,\mu$)-space. Nevertheless, our analysis shows quite consistent patterns in the dimensionless coordinates and provide an opportunity for their physical explanation.

  \item The similarity of SSB patterns in ($\xi,\,\mu$)-space for missions Jason-2 and Jason-3  for different data sets (years 2009 and 2018) proved robustness of the corresponding dependencies. At the same time, their dissimilarity with mission Jason-1 pointed out the problem of instrumentation and calibration (retracking) effects on the resulting SSB.

\item The theoretical consideration of sect.2 found the expected linear dependence of SSB on $\mu$: skewness bias (SB) of a random wave field is proportional to wave steepness like in the case of the classic stationary Stokes wave.   Previous  attempts to quantify such dependence as a contribution of one of the constituents of SSB, the skewness bias (SB), have been successful \citep{GommenBias2003}. Nevertheless, our theoretical result has not been supported by the data analysis. The most likely explanation is that all the effects contributing to SSB (e.g. EMB, SB, TB) are closely interrelated  and hardly discriminated within the parametric models.

\end{enumerate}

\section*{Acknowledgements}
\noindent Open access data of the portal AVISO  (http://www.aviso.altimetry.fr/en/home.html) has been used in this work. The data analysis (sect.~3) has been supported by the Russian Foundation for Basic Research $\sharp$19-05-00147А. Theoretical study of sect.~2 has been carried out under the Russian Science Foundation grant `Turbulence and coherent structures in the integrable and nonintegrable systems' $\sharp$19-72-30028 and Migo Group (http://migogroup.ru).
The authors are thankful to reviewers of the manuscript for their helpful critics.

%% The Appendices part is started with the command \appendix;
%% appendix sections are then done as normal sections
%% \appendix

%% \section{}
%% \label{}

%% If you have bibdatabase file and want bibtex to generate the
%% bibi
% \bibliographystyle{elsarticle-harv}
%\bibliography{bib_bsi_eng}
%%  \bibliography{<your bibdatabase>}

%% else use the following coding to input the bibitems directly in the
%% TeX file.

\end{document}